\documentclass[iop]{emulateapj}
\begin{document}

\received{}
\accepted{}

\slugcomment{Printed \today}

\title{
The counterjet of HH~30: new light on its binary driving source
}

\shorttitle{The Counterjet of HH~30}

\author{%
Robert Estalella\altaffilmark{1},
Rosario L\'opez\altaffilmark{1},
Guillem Anglada\altaffilmark{2}, 
Gabriel G\'omez\altaffilmark{3,4},\\
Angels Riera\altaffilmark{5,1},
and
Carlos Carrasco-Gonz\'alez\altaffilmark{2,6}
}
\altaffiltext{1}{%
Departament d'Astronomia i Meteorologia, 
Institut de Ciencies del Cosmos (ICC),
Universitat de Barcelona (IEEC-UB),
Mart\'{\i} i Franqu\`es 1, E-08028 Barcelona, Spain;
robert.estalella@am.ub.es, rosario.lopez@am.ub.es
}
\altaffiltext{2}{%
Instituto de Astrof\'{\i}sica de Andaluc\'{\i}a, CSIC,
Glorieta de la Astronom\'{\i}a s/n, E-18008 Granada, Spain;
guillem@iaa.es
}
\altaffiltext{3}{%
Instituto de Astrof\'{\i}sica de Canarias, 
E-38200 La Laguna, Tenerife, Spain; 
ggv@iac.es 
}
\altaffiltext{4}{%
GTC; GRANTECAN S.A. (CALP), 
E-38712 Bre\~na Baja, La Palma, Spain; 
gabriel.gomez@gtc.iac.es
}
\altaffiltext{5}{%
Departament de F\'{\i}sica i Enginyeria Nuclear, 
Escola Universit\`aria d'Enginyeria T\`ecnica Industrial de Barcelona, 
Universitat Polit\`ecnica de Catalunya, 
Comte d'Urgell 187, E-08036 Barcelona, Spain;
angels.riera@upc.edu
}
\altaffiltext{6}{%
Max-Planck-Institut f\"ur Radioastronomie,
Auf dem H\"ugel 69, 53121 Bonn, Germany 
}


\begin{abstract}

We present new [SII] images of the HH 30 jet and counterjet observed
in 2006, 2007, and 2010 that, combined with previous data, allowed us to
measure with improved accuracy the positions and proper motions of the jet and
counterjet knots. Our results show that the motion of the knots is essentially
ballistic, with the exception of the farthest knots, which trace the large
scale ``C''-shape bending of the jet. 
The observed bending of the jet can be produced by a relative motion of the HH
30 star with respect to its surrounding environment, caused either by a possible
proper motion of the HH 30 star, or by the entrainment of environment gas by the
red lobe of the nearby L1551-IRS 5 outflow. Alternatively, the bending can be
produced by the stellar wind from a nearby CTTS, identified in the 2MASS catalog
as J04314418+181047. 
The proper motion velocities of the knots of the
counterjet show more variations than those of the jet. In particular, we
identify two knots of the counterjet that have the same kinematic age but
whose velocities differ by almost a factor of two. Thus, it appears from our 
observations that counterjet knots launched simultaneously can be ejected with
very different velocities. We confirm that the observed wiggling of the jet
and counterjet arises from the orbital motion of the jet source in a binary 
system. Precession, if present at all, is of secondary importance in shaping
the jet. We derive an orbital period $\tau_o=114\pm2$ yr and a mass function
$m\mu_c^3=0.014\pm0.006$ $M_\odot$. 
For a mass of the system of $m=0.45\pm0.04$ $M_\odot$ (the value inferred from
observations of the CO kinematics of the disk) we obtain a mass
$m_j=0.31\pm0.04$ $M_\odot$ for the jet source, a mass $m_c=0.14\pm0.03$
$M_\odot$ for the companion, and a binary separation of $a=18.0\pm0.6$ AU. This
binary separation coincides with the value required to account for the size of
the inner hole observed in the disk, which has been attributed to tidal
truncation in a binary system.

\end{abstract}


\section{Introduction}

The Herbig-Haro (HH) object 30 \citep{mun83}, located in the northeastern part
of the L1551 dark cloud, lies at a distance of 140 pc \citep{ken94}. The HH 30
outflow is considered a prototypical jet/disk system. It presents a clear
jet/counterjet structure, which has been described by \citet{mun87, mun88} and
\citet{gra90}. The HH 30 exciting source is an optically invisible star
\citep{vrb85} highly extinguished by an edge-on disk \citep{bur96, sta99}, which
extends up to a radius of $\sim250$ AU perpendicularly to the jet, and divides
the surrounding reflection nebulosity into two lobes. \citet{ken98} propose a
spectral type M0 for the HH 30 star, and \citet{cot01} estimate a bolometric
luminosity of 0.2--0.9 $L_\odot$. \citet{lop95, lop96} propose that a number of
knots located to the northeast of the HH 30 object are also part of the same
flow, resulting in a total angular size of $\sim500''$ ($\sim0.35$ pc) for the
whole outflow. Several studies have explored 
the spatial morphology along and across the jet axis \citep{mun91, ray96}, 
the 3-D structure \citep{esq07,dec10}, 
line ratios \citep{mun90, bac99, har07},
or radial velocities \citep{rag97} of the HH 30 flow.

In addition, some studies have been carried out studying the variability of the
reflection nebulosity around HH 30. \citet{wat07} find
variations of the lateral asymmetry of the nebula and counternebula, but find no
convincing evidence for any period. However, \citet{dur09} find a periodic
modulation of the polarization of the nebula, with a period of 7.5 days, which
is interpreted as produced by asymmetric accretion hot spots on the star, or
orbiting clumps or voids in the disk.

Proper motions of a few knots of the HH 30 flow were determined by 
\citet{mun90}, \citet{bur96}, and \citet{lop96}. A thorough study of the 
proper motions of all the knots of the HH 30 jet and the closest knots 
of the counterjet was made by \citet{ang07}. The main result of this study 
is that the overall structure of the HH 30 jet can be well fitted by the 
shape of a wiggling ballistic jet, arising either by the orbital motion of 
the jet source or by precession of the jet axis because of the tidal 
effects of a companion. In the first case 
the binary separation is expected to be 9--18 AU, 
while in the case of precession 
the binary separation is $<1$ AU. 
Given that the radius of the flared disk observed in the HST images is 250 AU,
the conclusion is that this disk appears to be a circumbinary disk rather than
a circumstellar disk, contrary to what was initially thought. This circumbinary
disk is unlikely to have a relevant role in the jet collimation.

\citet{pet06} carry out PdBI observations of the dust continuum and CO line
emission revealing the presence of an asymmetric molecular outflow and a disk
in Keplerian rotation, deriving a central stellar mass of $0.45\pm0.04$ 
$M_\odot$. 
More recently, \citet{gui08} carry out CO and dust  continuum observations of
the circumbinary disk that reveal an inner cavity with a radius of $37\pm4$ AU,
and \citet{mad12}, throught a detailed modeling, conclude that the
disk has an inner depletion zone with a similar radius os $45\pm5$ AU
radius. 
\citet{gui08} explain the size of the  inner hole as a result of tidal forces
induced by a binary with a  separation of $18\pm2$ AU, supporting the binary
interpretation proposed  by \citet{ang07}. 

In this paper we present new [SII] images of the HH 30 jet and counterjet
carried out in 2006, 2007, and 2010, which allowed us to measure the position
and proper motion of the jet and counterjet knots, and to constrain the
parameters of the binary system. In \S 2 we describe the observations; in \S 3
we describe the procedure used for the proper motions determination; in \S 4 we
study the large-scale jet bending, and its possible origin; in \S 5 we analyze
the jet wiggling and the constraints on the physical parameters of the binary
system at the core of HH 30; finally, in \S 6 we discuss the results obtained
and in \S 7 we give our conclusions.

\section{Observations}

The CCD observations used in this paper to determine the proper motions of 
the HH~30 jet/counterjet system are listed in Table \ref{log}. All the 
images were obtained through [SII] narrow-band filters, which included the 
$\lambda\lambda$6716, 6731 \AA\ emission lines. The details on the setup 
configuration, acquisition and treatment of the 2.5~m Isaac Newton 
Telescope (INT) image are given in \citet{lop95}. All the images obtained 
at the 2.6~m Nordic Optical Telescope (NOT) were obtained using the same 
setup configuration (i.e.\ the Andaluc\'ia Faint Object Spectrograph and 
Camera, ALFOSC, and the [SII] filter centered on $\lambda=6724$ \AA\ and 
bandpass $\Delta\lambda=50$ \AA). More details on the acquisition and 
treatment of the NOT observations can be found in \citet{ang07}. 
Finally, the 2010 image was obtained at the 4.2~m William Herschel 
Telescope (WHT), using the ACAM camera on the Cassegrain focus, giving a 
field of view of $8'$ with a spatial scale of $\sim 0\farcs25$ 
pixel$^{-1}$. A narrow-band [SII] filter, centered on $\lambda=6727$ \AA\ 
and bandpass $\Delta\lambda=48$ \AA\ was used. In addition, a frame of a 
shorter exposure time of 3600 s was acquired through another narrow-band 
filter, centered on $\lambda=6645$ \AA\ and bandpass $\Delta\lambda=50$ 
\AA. This filter includes the nearby continuum, free of [SII] emission 
lines, useful to subtract the contribution of the continuum reflected emission
close to the HH~30 jet source from the line emission image. For each 
epoch, several frames, with a typical time exposure of 1800~s, were obtained
to complete the total integration times, listed in Table \ref{log}, of the
final deep images. The individual frames were processed using the standard
tasks of IRAF\footnote{%
IRAF is distributed by the National Optical Astronomy Observatories, 
which are operated by the Association of Universities for Research in 
Astronomy, Inc., under cooperative agreement with the National Science 
Foundation.} 
reduction package, which included bias subtraction and flat-field corrections,
using sky flats. In order to correct for misalignments, all the individual
frames of the same epoch were recentered using the position of field stars.
Then, the frames were median-averaged to obtain a final deep image for each of
the epochs listed in Table~\ref{log}. Images were not flux calibrated.

\begin{deluxetable}{lccc}
\tablecaption{[SII] images of the HH~30 jet/counterjet system}
\tablecolumns{4}
\tablehead{
&& 
Exp.\ time 
& \\
Epoch & 
Telescope/Instrument & 
(s) & 
Reference
}
\startdata
1993 Dec 15 & INT        &    14400 & 1 \\ 
1998 Nov 08 & NOT/ALFOSC & \phn9000 & 2 \\ 
1999 Nov 20 & NOT/ALFOSC &    14400 & 2 \\ 
2006 Jan 19 & NOT/ALFOSC &    14400 & 3 \\ 
2007 Nov 06 & NOT/ALFOSC &    10800 & 3 \\ 
2010 Dec 01 & WHT/ACAM   &    12000 & 3 
\enddata
\tablerefs{
(1) \citet{lop95}
(2) \citet{ang07}
(3) This work
}
\label{log}
\end{deluxetable}

The six final images were converted into a common reference system and rebinned
to the same pixel scale. The positions of seven field stars, common to all the
frames, were used to register the images. The GEOMAP and GEOTRAN tasks of IRAF
were applied to perform a linear transformation, with six free parameters that
take into account translation, rotation and magnification between different
frames. All the transformed frames have a pixel size equal to that of the last
epoch WHT image that was taken as the reference image. 

Astrometric calibration of the images transformed to the common reference system
was made by using the coordinates of ten field stars, well distributed on the
observed field. The coordinates were obtained after identifying the stars from
the 2MASS All Sky Catalogue. The high optical extinction of the region made it
impossible to find such a grid of reference stars from the USNO-B1.0 Catalogue.
The typical rms of the transformation was $\leq0\farcs1$ in both coordinates.
The pixel size was found to be $0\farcs2528$ pixel$^{-1}$.

\section{Proper motions}
\label{pmotion}

\begin{deluxetable*}{lrrrrrrrr}
\tabletypesize{\small}
\tablecaption{Positions and Proper Motions of Knots in the HH~30 Jet
}
\tablecolumns{9}
\tablehead{
&
\colhead{$x$\tablenotemark{a}}&
\colhead{$y$\tablenotemark{a}}&
\colhead{$\mu_x$}&
\colhead{$\mu_y$}&
\colhead{$\epsilon_x$\tablenotemark{b}}&
\colhead{$\epsilon_y$\tablenotemark{b}}&
\colhead{$v_t$\tablenotemark{c}}&
\colhead{P.A.\tablenotemark{d}}
\\
\colhead{Knot}&
\colhead{($''$)}& 
\colhead{($''$)}& 
\colhead{($''$ yr$^{-1}$)}&
\colhead{($''$ yr$^{-1}$)}&
\colhead{($''$)}& 
\colhead{($''$)}& 
\colhead{(km s$^{-1}$)}&
\colhead{(deg)}
}
\startdata
A1 & $-0.04$& $2.69 $& $-0.005\pm0.007$& $0.153\pm0.010$& $0.06$& $0.08$& $101.5\pm\phn6.4$& $ 1.9\pm2.6$\\
A2 & $ 0.08$& $4.12 $& $ 0.007\pm0.007$& $0.150\pm0.008$& $0.08$& $0.08$& $ 99.5\pm\phn5.1$& $-2.6\pm2.8$\\
A3 & $ 0.09$& $6.85 $& $ 0.003\pm0.003$& $0.179\pm0.008$& $0.03$& $0.08$& $118.7\pm\phn5.1$& $-1.0\pm0.9$\\
B1 & $ 0.12$& $11.04$& $-0.005\pm0.003$& $0.097\pm0.006$& $0.03$& $0.06$& $ 64.5\pm\phn3.9$& $ 3.0\pm1.9$\\
B2 & $-0.10$& $13.49$& $-0.005\pm0.002$& $0.193\pm0.007$& $0.03$& $0.10$& $127.9\pm\phn4.5$& $ 1.5\pm0.7$\\
B3 & $-0.25$& $14.36$& $ 0.008\pm0.008$& $0.190\pm0.010$& $0.08$& $0.10$& $126.4\pm\phn6.8$& $-2.4\pm2.5$\\
C	 & $-0.30$& $17.12$& $-0.005\pm0.003$& $0.177\pm0.017$& $0.04$& $0.24$& $117.3\pm   11.4$& $ 1.6\pm1.0$\\
D1 & $ 0.05$& $21.19$& $ 0.009\pm0.007$& $0.319\pm0.028$& $0.07$& $0.29$& $212.1\pm   18.6$& $-1.5\pm1.3$\\
D2 & $ 0.06$& $24.31$& $-0.005\pm0.007$& $0.382\pm0.024$& $0.10$& $0.33$& $253.8\pm   15.8$& $ 0.7\pm1.1$\\
D3 & $ 0.09$& $27.17$& $-0.008\pm0.005$& $0.265\pm0.098$& $0.02$& $0.34$& $175.9\pm   64.8$& $ 1.8\pm1.2$\\
D4 & $-0.33$& $29.64$&          \nodata&         \nodata&\nodata&\nodata&            \nodata&      \nodata\\
E1 & $-0.60$& $35.81$& $-0.005\pm0.002$& $0.149\pm0.007$& $0.03$& $0.10$& $ 98.8\pm\phn4.7$& $ 2.0\pm0.8$\\
E2 & $ 0.60$& $41.05$& $-0.001\pm0.009$& $0.099\pm0.012$& $0.12$& $0.17$& $ 65.9\pm\phn7.9$& $ 0.8\pm5.0$\\
E3b& $-0.55$& $48.03$& $-0.007\pm0.009$& $0.199\pm0.004$& $0.08$& $0.04$& $131.9\pm\phn2.9$& $ 1.9\pm2.5$\\
E4 & $-0.46$& $52.04$& $-0.020\pm0.007$& $0.143\pm0.003$& $0.10$& $0.05$& $ 95.8\pm\phn2.3$& $ 7.8\pm2.7$
\enddata
\tablenotetext{a}
{Positions in the 2010 image, except for knot B3, for which the position is that of
the 2007 image. The $y$ axis is along the jet, at a position angle (eastwards from
north) of $31\fdg6$. The $(0,0)$ position is that of the brightest knot, A0.}
\tablenotetext{b}
{Rms residual of the knot positions in the proper motion fit, 
$\epsilon=\sigma\sqrt{1-r^2}$, where $\sigma$ is the standard deviation 
and $r$ the correlation coefficient.}
\tablenotetext{c}
{Proper motion velocity, assuming a distance of 140 pc.}
\tablenotetext{d}
{Position angle with respect to the $y$ direction.}
\label{pmjet}
\end{deluxetable*}

\begin{deluxetable*}{lrrrrrrrr}
\tabletypesize{\small}
\tablecaption{Positions and Proper Motions of Knots in the HH~30 Counterjet
}
\tablecolumns{9}
\tablehead{
&
\colhead{$x$\tablenotemark{a}}&
\colhead{$y$\tablenotemark{a}}&
\colhead{$\mu_x$}&
\colhead{$\mu_y$}&
\colhead{$\epsilon_x$\tablenotemark{b}}&
\colhead{$\epsilon_y$\tablenotemark{b}}&
\colhead{$v_t$\tablenotemark{c}}&
\colhead{P.A.\tablenotemark{d}}
\\
\colhead{Knot}&
\colhead{($''$)}& 
\colhead{($''$)}& 
\colhead{($''$ yr$^{-1}$)}&
\colhead{($''$ yr$^{-1}$)}&
\colhead{($''$)}& 
\colhead{($''$)}& 
\colhead{(km s$^{-1}$)}&
\colhead{(deg)}
}
\startdata
Z1 & $ 0.05$& $  -5.61$& $ 0.000\pm0.010$& $-0.278\pm0.012$& $0.08$& $0.13$& $184.8\pm\phn8.3$& $-179.9\pm\phn2.1$\\
Z2 & $-0.05$& $  -9.55$& $ 0.011\pm0.001$& $-0.462\pm0.003$& $0.01$& $0.03$& $307.0\pm\phn2.2$& $-178.7\pm\phn0.1$\\
Z3 & $ 0.16$& $ -11.71$& $ 0.065\pm0.020$& $-0.223\pm0.026$& $0.07$& $0.09$& $154.4\pm   17.1$& $-163.7\pm\phn5.1$\\
Z4 & $-0.23$& $ -13.43$& $-0.022\pm0.022$& $-0.166\pm0.055$& $0.08$& $0.19$& $111.0\pm   36.0$& $ 172.5\pm\phn8.0$\\
Z5a& $ 0.06$& $ -17.41$& $ 0.046\pm0.046$& $-0.227\pm0.003$& $0.16$& $0.01$& $153.8\pm\phn6.4$& $-168.6\pm   11.3$\\
Z5b& $-0.10$& $ -18.26$& $-0.027\pm0.006$& $-0.316\pm0.003$& $0.02$& $0.01$& $210.7\pm\phn1.9$& $ 175.1\pm\phn1.0$\\
Z6 & $ 0.16$& $ -23.65$& $ 0.012\pm0.016$& $-0.608\pm0.084$& $0.05$& $0.29$& $403.3\pm   55.5$& $-178.9\pm\phn1.5$\\
J	 & $ 0.32$& $ -78.30$& $ 0.029\pm0.013$& $ 0.066\pm0.060$& $0.17$& $0.78$& $ 48.0\pm   36.9$& $ -23.6\pm   21.5$\\
K1 & $ 1.56$& $ -94.35$& $ 0.064\pm0.016$& $-0.174\pm0.057$& $0.20$& $0.73$& $122.9\pm   35.5$& $-159.8\pm\phn7.6$\\
K2 & $ 0.89$& $ -99.02$& $ 0.036\pm0.012$& $-0.148\pm0.032$& $0.16$& $0.42$& $101.3\pm   21.0$& $-166.4\pm\phn5.4$\\
L	 & $ 4.09$& $-149.00$& $ 0.004\pm0.012$& $-0.249\pm0.025$& $0.16$& $0.32$& $165.0\pm   16.6$& $-179.2\pm\phn2.8$\\
M	 & $ 7.23$& $-176.61$& $ 0.011\pm0.008$& $-0.325\pm0.015$& $0.10$& $0.20$& $215.9\pm   10.2$& $-178.0\pm\phn1.4$\\
N	 & $11.59$& $-191.00$& $ 0.076\pm0.003$& $-0.323\pm0.015$& $0.04$& $0.19$& $220.1\pm\phn9.5$& $-166.7\pm\phn0.8$
\enddata
\tablenotetext{a}
{Positions in the 2010 image. The $y$ axis is along the jet, 
at a position angle of $31\fdg6$. The $(0,0)$ position is that of the brightest knot, A0.}
\tablenotetext{b}
{Rms residual of the knot positions in the proper motion fit, 
$\epsilon=\sigma\sqrt{1-r^2}$, where $\sigma$ is the standard deviation 
and $r$ the correlation coefficient.}
\tablenotetext{c}
{Proper motion velocity, assuming a distance of 140 pc.}
\tablenotetext{d}
{Position angle with respect to the $y$ direction.}
\label{pmcjet}
\end{deluxetable*}

The images used for proper motion determination were rotated by an angle of 
$31\fdg6$, the position angle of the jet axis \citep{ang07}, so that the 
$y$ axis is along the jet axis and the $x$-axis is perpendicular, with values 
increasing from left to right.
In order to improve the signal-to-noise ratio, the images were smoothed with a
Gaussian with a FWHM of 3 pixels. 
The smoothed images were used for the farther away knots of the jet and counterjet
(B to E, and J to N), while the full resolution images were used for the
stronger, closer to the origin knots (A and Z). 
This nomenclature for the knots is based on that of \citet{lop95} and
\citet{ang07}. Knots Z3 to Z6 of the counterjet, detected in our 2006, 2007, and
2010 observations, have been identified for the first time.
Positions of the knots for
each epoch were measured with respect to the position of the brightest knot A0,
whose position was set to $(x,y)=(0,0)$. The position of the knots was
determined from a parabolic fit to the intensity of the $5\times5$ pixels
centered on the pixel with peak emission. Then, the proper motion in $x$ and
$y$ directions of each knot, $\mu_x$ and $\mu_y$, was determined from a linear
regression fit to their positions in the different epoch images.

\begin{figure*}
\plotone{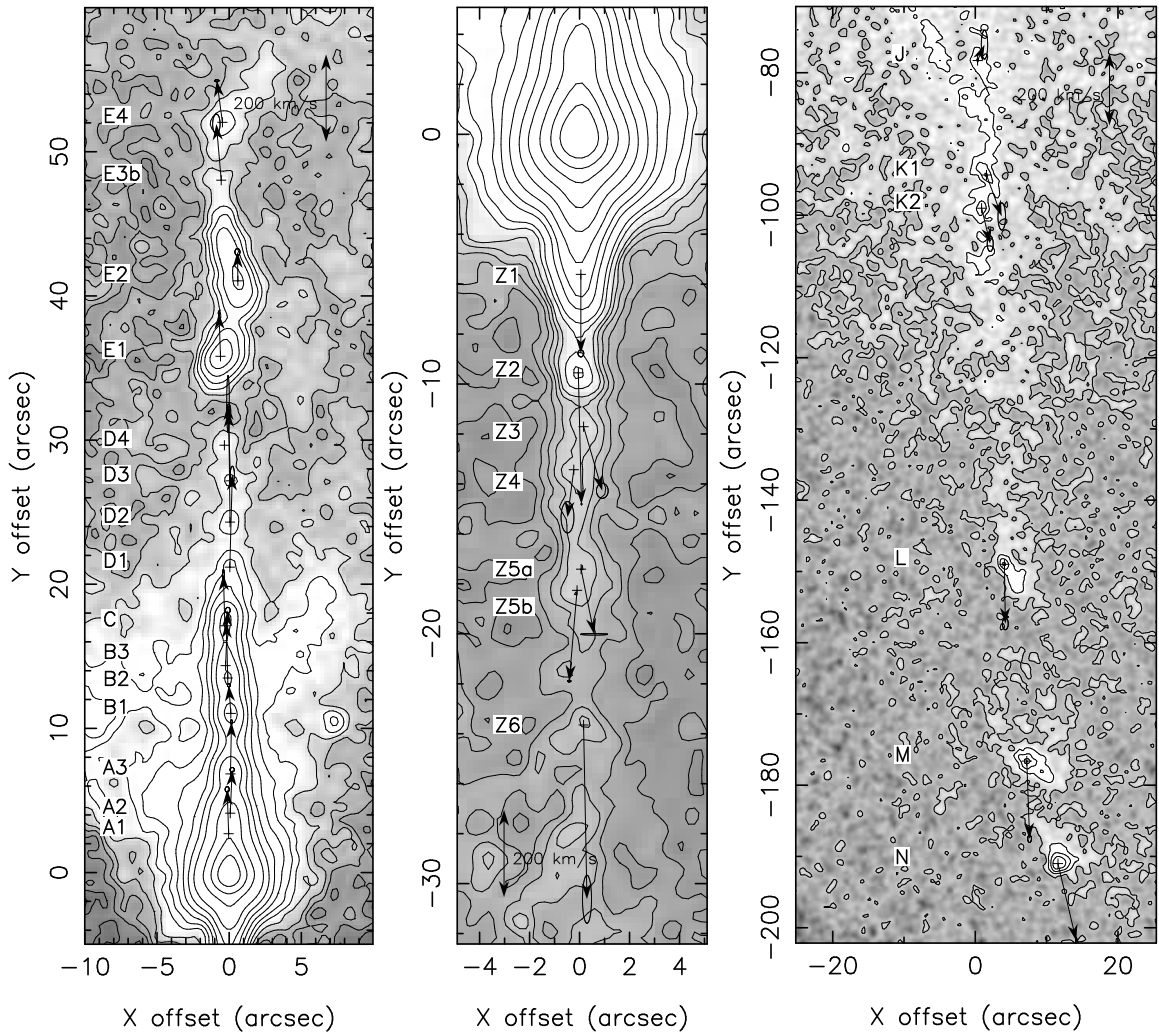}
\caption{2010 [SII] narrow-band image of the jet and counterjet of HH 30, 
smoothed with a 3-pixel Gaussian, showing the proper motion velocities of the 
knots A to E (\emph{left}); 
knots Z (\emph{center}); and
knots J to N (\emph{right}).
The $y$ axis is along the jet axis, at a position angle of $31\fdg6$.
}
\label{fpmae}
\end{figure*}

The 1993 image has poorer seeing than the others, and could not be used for
knots close to the HH 30 star, i.e.\ knots A and Z. The 1989 and 1999 images do
not include the counterjet region for distances greater than $5\farcs6$, and
could not be used neither for the counterjet knots Z3 to Z6, nor for J to N.
The results are given in Tables \ref{pmjet} and \ref{pmcjet}, and shown in
Fig.\ \ref{fpmae}, 
where the knot positions and proper motions are shown superimposed on the
smoothed 2010 image. 

\begin{figure*}
\plotone{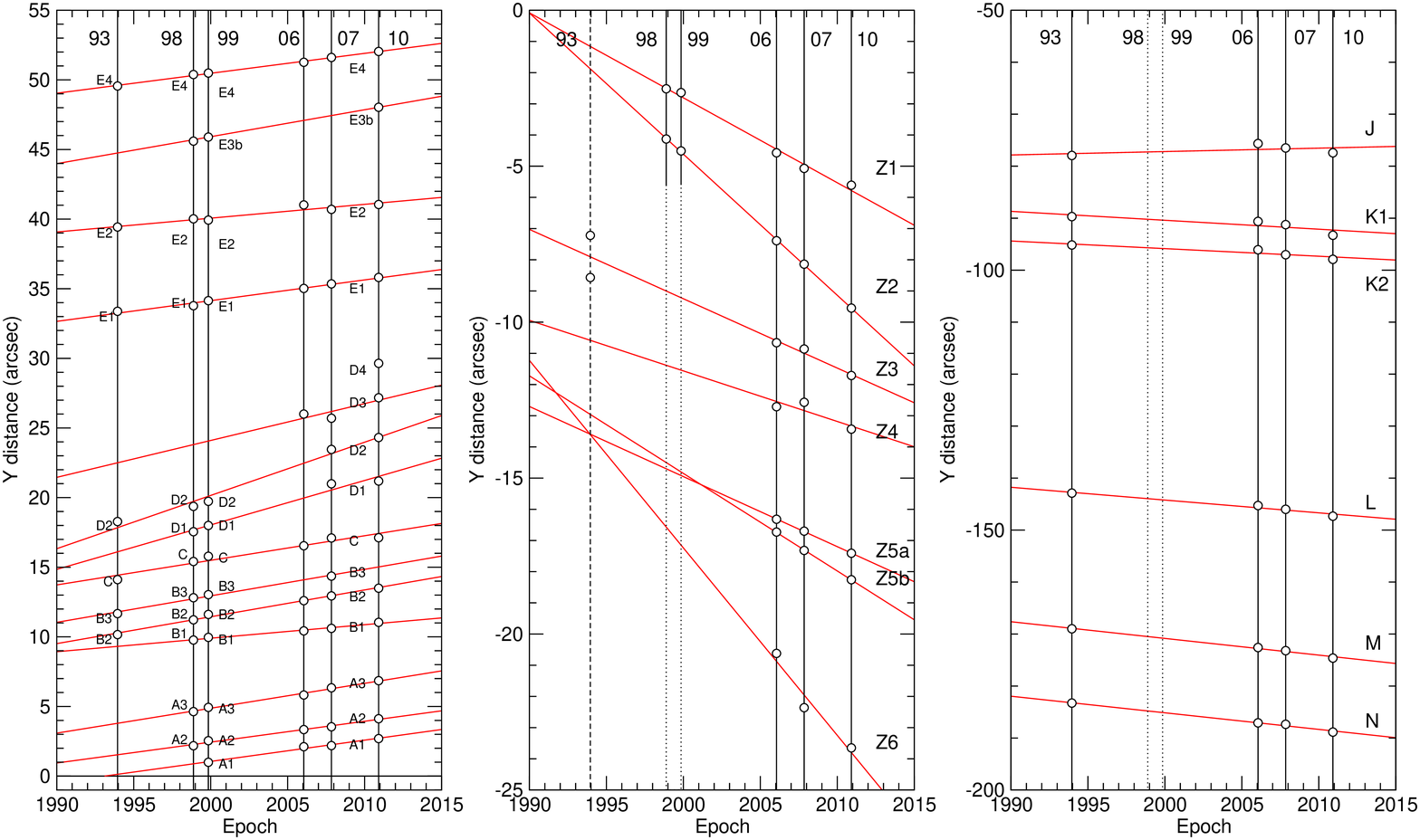}
\caption{Diagram showing the $y$ position of the knots of the jet and
counterjet, as a function of time:
knots A to E (\emph{left}); 
knots Z (\emph{center}); and
knots J to N (\emph{right}).
The red lines show the linear fits used to derive the proper motion velocity in 
the $y$ direction. The positions of 1993 have not been used for the proper 
motion determination of knots Z. The 1998 and 1999 images only reach up to 
$y=-5\farcs6$ from the HH 30 star in the counterjet.
}
\label{fyae}
\end{figure*}

The errors assigned to the proper motions, in both $x$ and $y$, are the 
formal errors of the slope of the linear regression fits for each knot. 
These errors are shown as ellipses at the end of the arrows shown in 
Fig.\ \ref{fpmae}. 
The errors appear to be rather small as a consequence of the wide span of time
used for calculating the proper motions, and that 6 epochs were used for most 
knots. The quality of the linear regression fits for each knot is indicated by
the small values of the residuals in $x$, $\epsilon_x$, and $y$, $\epsilon_y$,
(see Tables \ref{pmjet} and \ref{pmcjet}) and is illustrated in 
Fig.\ \ref{fyae}, 
where the measured and fitted values of the $y$ position as a function of time
for each knot, are shown. In all the cases, the $y$ positions are well fitted
by the proper motion velocity obtained, showing that the motion of the knots is
ballistic. 

All the knots show proper motions roughly in the direction of the jet 
axis, with the exception of knot J of the counterjet, which has been 
measured to be nearly stationary. This knot is located inside a 
reflection nebulosity southwest of the HH 30 star (spreading from 
$y\simeq-25''$ to $-100''$; see Fig.\ \ref{fpmae}), 
and is most probably not a counterjet knot, but a 
feature of the reflection nebulosity. We will not consider this knot in 
the following.

It is worth noting that the proper motions $\mu_y$ obtained here are 
significantly lower than those obtained in \citet{ang07} from two observations
with nearly one year of interval, in 1998 and 1999. On the average, the proper
motion velocities for knots A to E of the jet are $0\farcs11$ yr$^{-1}$ (or 75
km s$^{-1}$) higher in \citet{ang07} than the present values. For the
counterjet, the absolute value of the proper motion velocities for knots Z1 and
Z2 are $0\farcs13$ yr$^{-1}$ (or 85 km s$^{-1}$) lower in \citet{ang07} than the
present values. 
Since the proper motions are measured with respect to the brightest knot A0, the
discrepancy can be explained by a shift in the position of A0 along the axis of
the jet of $\Delta{y}\simeq-0\farcs12$ in the 1999 image with respect to that of
1998. Such a shift in position can be attributed to the observed variations in
the brightness of the diffuse light of the HH 30 star \citep{wat07, dur09}.
However, this systematic error in the determination of proper motions is not
expected to affect measurements that span over a large  number of years, like
those presented in the present work.

\begin{figure}
\plotone{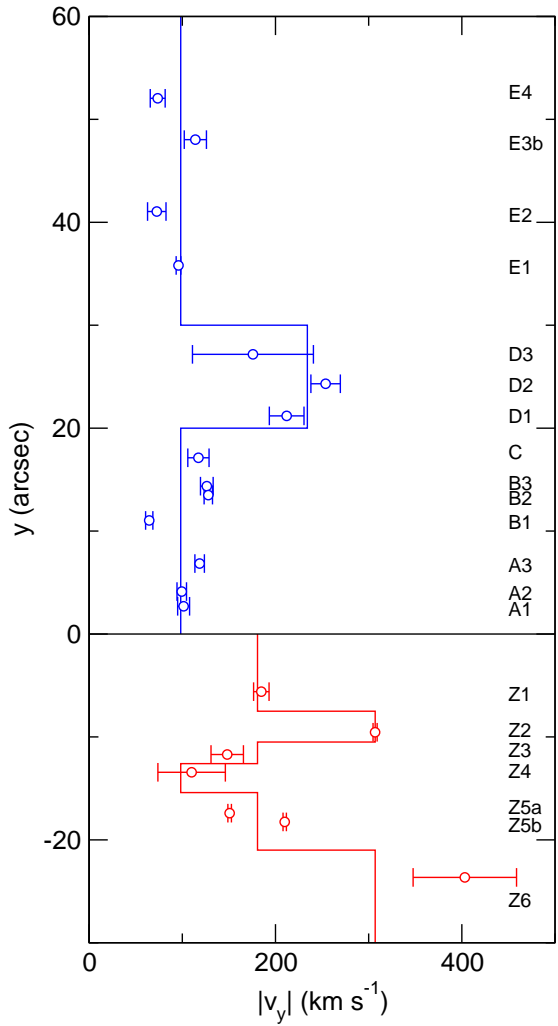}
\caption{Proper motion velocity in the $y$ direction (absolute value), $|v_y|$,
versus distance, $y$, for the knots in the HH~30 jet (blue) and counterjet
(red).
The stepped lines indicate the velocities adopted for the knots of the jet
(blue) and counterjet (red). These velocities are the error-weighted averages of
the velocity of the knots in each step: 
$v_j=98$ km s$^{-1}$ (A, B, C, E, Z4); 
$v_j=181$ km s$^{-1}$ (Z1, Z3, Z5); 
$v_j=234$ km s$^{-1}$ (D); and 
$v_j=307$ km s$^{-1}$ (Z2, Z6). 
}
\label{fvyy}
\end{figure}

In Fig.\ \ref{fvyy} we plot the proper motion velocity in the $y$ direction, in
absolute value, $|v_y|$, as a function of distance $y$. As can be seen, the jet
and counterjet velocities are quite different. 
For the jet knots, most have velocities $\sim100$ km s$^{-1}$,
except knots D, with distances from the HH 30 star between $20''$ and $30''$,
which have a higher velocity, $\sim240$ km s$^{-1}$. 
For the counterjet, the velocities are more irregular. There are knots with
$|v_y|\simeq100$ km s$^{-1}$ (Z4), with 
$|v_y|\simeq180$ km s$^{-1}$ (Z1, Z3, Z5a, Z5b), and with 
$|v_y|>300$ km s$^{-1}$ (Z2, Z6). 
On the average, the proper motion
velocity of the jet knots is 115 km s$^{-1}$, while for the counterjet knots the
average is 193 km s$^{-1}$. The difference is also found
in the velocity dispersion, which is 54 km s$^{-1}$ for the jet, and 82 km
s$^{-1}$ for the counterjet. Thus, the counterjet knots
have, on average, higher velocities than the jet knots, and show more
variation from knot to knot. 

Let us now consider the pair of knots of the counterjet Z1 and Z2. The 
velocity of Z2 ($307\pm2$ km s$^{-1}$) is much higher than that of Z1 
($185\pm8$ km s$^{-1}$). This pair of knots, in spite of being very close 
to the HH 30 star, have well determined proper motions, since they have 
been observed at five epochs, spanning 17 years. In addition, the proper 
motions obtained by fitting only the new observations in 2006, 2007, and 
2010, coincide with those obtained when the old observations of 1998 and 
1999 are included in the fit. The positions of the two knots Z1 and Z2, 
when extrapolated back in time, show that they were launched 
simultaneously, at the end of 1989 (see Fig.\ \ref{fyae}). A similar case 
is found for the pair of knots Z5 and Z6 of the counterjet, but in this 
case the proper motion velocities are not so well constrained as in the 
case of Z1 and Z2. Thus, it appears that the counterjet of HH 30 is able 
to launch simultaneously pairs of knots with very different velocities.

\begin{deluxetable}{lrr}
\tablecaption{Launch properties of knots in the HH~30 jet and counterjet}
\tablecolumns{3}
\tablehead{
&
\colhead{$t_\mathrm{kin}$\tablenotemark{a}}&  
\colhead{$x_\mathrm{lch}$\tablenotemark{b}}
\\
\colhead{Knot}&
\colhead{(yr)}& 
\colhead{(arcsec)}
}
\startdata
E4 & $364.1\pm\phn1.3$& $  6.7\pm\phn2.5$\\
E3b& $248.3\pm\phn8.4$& $  3.7\pm\phn3.3$\\
E2 & $413.7\pm\phn4.9$& $  1.1\pm\phn3.6$\\
E1 & $240.8\pm\phn1.7$& $  0.7\pm\phn0.5$\\
D4 & \nodata         &  \nodata          \\
D3 & $102.6\pm10.0   $& $  1.0\pm\phn0.6$\\
D2 & $ 63.6\pm\phn1.5$& $  0.4\pm\phn0.5$\\
D1 & $ 66.3\pm\phn1.9$& $ -0.5\pm\phn0.5$\\
C  & $ 96.9\pm\phn1.7$& $  0.2\pm\phn0.3$\\
B3 & $ 75.4\pm\phn0.8$& $ -0.8\pm\phn0.6$\\
B2 & $ 70.0\pm\phn0.5$& $  0.2\pm\phn0.2$\\
B1 & $113.6\pm\phn0.7$& $  0.7\pm\phn0.4$\\
A3 & $ 38.3\pm\phn0.3$& $  0.0\pm\phn0.1$\\
A2 & $ 27.5\pm\phn0.2$& $ -0.1\pm\phn0.2$\\
A1 & $ 17.6\pm\phn0.2$& $  0.1\pm\phn0.1$\\
Z1 & $ 20.1\pm\phn0.3$& $  0.0\pm\phn0.2$\\
Z2 & $ 20.6\pm\phn0.1$& $ -0.3\pm\phn0.1$\\
Z3 & $ 52.4\pm\phn1.4$& $ -3.3\pm\phn1.1$\\
Z4 & $ 81.0\pm\phn4.4$& $  1.6\pm\phn1.9$\\
Z5a& $ 76.6\pm\phn0.2$& $ -3.4\pm\phn3.6$\\
Z5b& $ 57.7\pm\phn0.2$& $  1.5\pm\phn0.3$\\
Z6 & $ 38.9\pm\phn3.3$& $ -0.3\pm\phn0.6$\\
K1 & $543.0\pm30.8   $& $-33.1\pm14.2   $\\
K2 & $667.6\pm21.7   $& $-23.0\pm\phn9.8$\\
L  & $599.2\pm15.0   $& $  1.9\pm\phn7.3$\\
M  & $543.2\pm\phn8.3$& $  1.0\pm\phn4.2$\\
N  & $592.0\pm\phn8.7$& $-33.6\pm\phn2.7$
\enddata
\tablenotetext{a}
{Kinematic age defined as $t_\mathrm{kin}=y/\mu_y$.}
\tablenotetext{b}
{Offset position at launch, defined as $x_\mathrm{lch}=x-\mu_x t_\mathrm{kin}$.}
\label{tlaunch}
\end{deluxetable}

\begin{figure}
\plotone{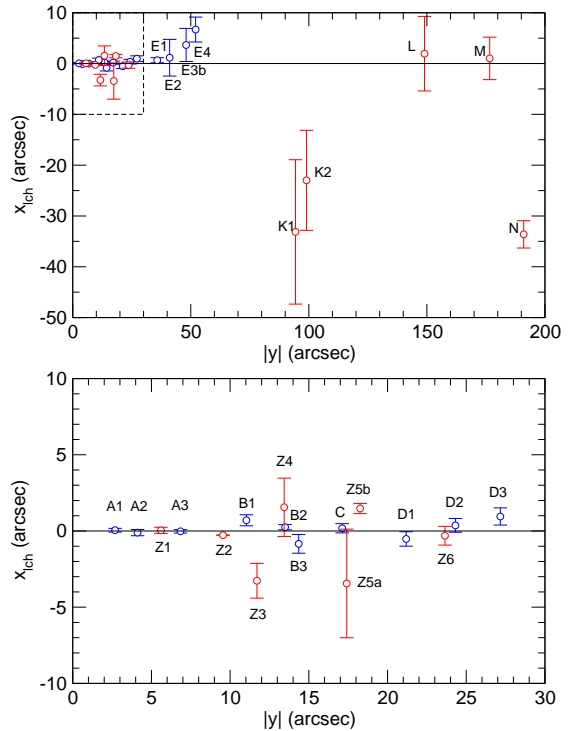}
\caption{Position in the $x$ axis at the epoch of launch assuming ballistic 
motions, $x_\mathrm{lch}=x-y\,\mu_x/\mu_y$, versus $|y|$ for the knots in the 
HH~30 jet (blue) and counterjet (red).  
\emph{Top:} All knots. The dashed line shows
the region that appears enlarged in the bottom panel. 
\emph{Bottom:} Knots close to the origin.
}
\label{fxlt}
\end{figure}

In order to investigate the launching properties of all the knots, we 
calculated for each knot its kinematic age, 
$t_\mathrm{kin}\simeq y/\mu_y$, 
and the position at launch, 
$x_\mathrm{lch}\simeq x-\mu_x t_\mathrm{kin}$. 
The results obtained are  shown in Table \ref{tlaunch}. In Fig.\ \ref{fxlt} we
show the $x$  position at launch, $x_\mathrm{lch}$, as a function of distance to
the  origin, $|y|$. As can be seen, all knots, except the more distant knots, 
have values of $x_\mathrm{lch}$ consistent with zero. This is an  indication
that these knots are ballistic, moving along a straight line  after their
launching. For the more distant knots, K1, K2, and N, 
$x_\mathrm{lch}\simeq-30''$, indicating that they deviate from a straight  line
after their launching. As we will see in \S \ref{bending}, the more  distant
knots trace the large-scale ``C''-shape bending of the jet and  counterjet.

\section{Jet bending}
\label{bending}

\subsection{Bending by a plane-parallel side wind}

The bending of the HH 30 jet/counterjet, with a clear ``C'' shape, was reported
by \citet{lop95} and \citet{ang07}. Now, with the observations of the counterjet
presented in this work, we know the shape of the jet/counterjet for a length of
$\sim500''$, corresponding to $\sim0.35$ pc.

The bending of the HH 30 jet is suggestive of being entrained toward the 
north-west (increasing $x$). This could be due to a proper motion of the HH 30
star toward the south-east with respect to the ambient medium. \citet{can95}
estimate that a relative velocity between the HH 30 star and the surrounding
environment of $\sim2$ km s$^{-1}$ is sufficient for explaining the observed
bending of the HH 30 jet/counterjet. Such a relative velocity could be the
result of a proper motion of the HH 30 star of $\sim0\farcs003$ yr$^{-1}$,
clearly below our accuracy in the proper motion measurements, of the order of
$\sim0\farcs1$ yr$^{-1}$. An alternative is that the deflection is caused by
the powerful bipolar outflow whose exciting source, L1551-IRS5, is at a
distance of $\sim4'$ southwards of HH 30. The red lobe of the L1551-IRS5
molecular outflow is found to the south-east of the HH 30 star, and could be
responsible of entraining the surrounding environment with a velocity
of $\sim2$ km s$^{-1}$ and deflecting the jet and counterjet of HH 30.

A different explanation for the bending of the HH 30 jet and counterjet is a
deflection by an isotropic stellar wind blowing the HH 30 jet from the
south-east. In the following we will examine this possibility.

\subsection{Bending by an isotropic stellar wind}

\begin{figure}
\plotone{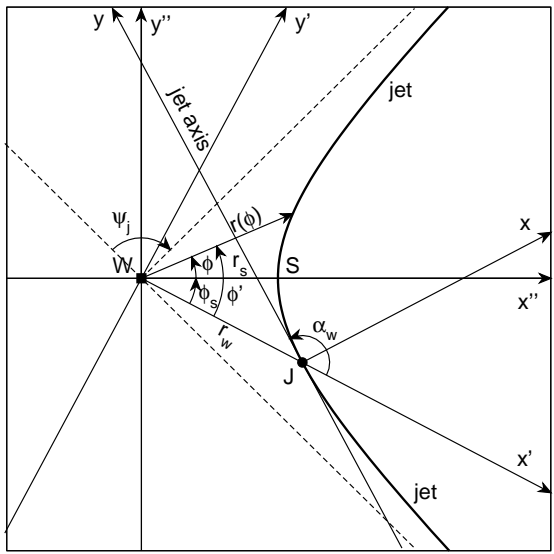}
\caption{Reference systems used for studying the jet and counterjet bending. 
The jet path is indicated by the thick continuum line. 
The jet source, $J$, is indicated with a small circle, while the wind source,
$W$, is indicated with a small square. 
The ``stagnation point'' is $S$. 
The distance from the side-wind source to $S$ is $r_s$, and to $J$ is $r_w$.  
The angle $\alpha_w$ is the angle between the $x'$ and $y$ axes, and
$\phi_s$ is the angle  with vertex at $W$, subtended by the jet source $J$ and
the stagnation point $S$.  
The reference system $(x',y')$ is rotated an angle $\phi_s$ with respect to
$(x'', y'')$. 
The reference system $(x, y)$ is shifted a distance $r_w$ and rotated an angle
$\alpha_w-\pi/2$ with respect to $(x', y')$. 
The two dashed lines indicate the asymptotes to the jet and counterjet.
The total deflection of the jet, $\psi_j$, is the angle between the two
asymptotes.}
\label{fig_raga09}
\end{figure}

We will use the model of jet deflection by an isotropic stellar wind, developed
by \citet{rag09a}, to fit the shape of the HH 30 jet/counterjet. The path of the
jet/counterjet in a reference system centered on the wind source (see Fig.\
\ref{fig_raga09}), is given in polar coordinates by
\begin{equation}\label{rphi}
r(\phi)=\frac{r_s}{\cos^\lambda(\phi/\lambda)}, \;\; 
-\lambda\frac{\pi}{2}\ < \phi < \lambda\frac{\pi}{2},
\end{equation}
where $r_s$ is the distance from the wind source to the ``stagnation
point'' (i.e.\ the point closest to the wind source) of the jet/counterjet
system, and 
\begin{equation}\label{elambda}
\lambda\equiv\frac{\epsilon}{1+\epsilon}, \;\; 0< \lambda< 1,
\end{equation}
with $\epsilon$ given by
\begin{equation}\label{eq_epsilon}
\epsilon=
2 \left(\frac{\dot{M}_j v_j}{\dot{M}_w v_w}\right)^{1/2} \frac{v_j}{c_j},
\end{equation}
where $\dot{M}_j$, $v_j$ and $c_j$ are the mass-loss rate, velocity and
isothermal sound speed of the jet, and $\dot{M}_w$, $v_w$ are the 
mass-loss
rate and velocity of the wind. The parameter $\lambda$ can be expressed in
terms of the total deflection angle of the jet/counterjet path, $\psi_j$ (see 
Fig.\ \ref{fig_raga09}),
\begin{equation}\label{psi}
\lambda= 1-\frac{\psi_j}{\pi}.
\end{equation}

Let us call $r_w$ the distance from the jet source to the wind source, and
$\alpha_w$ the angle between the jet axis ($y$ axis) and 
the direction from the wind source to the jet source ($x'$ axis). 
As shown by \citet{rag09a}, the angle $\phi_s$, with vertex at the wind
source and subtended by the jet source and the stagnation point (see Fig.\
\ref{fig_raga09}), is given by. 
\begin{equation}
\phi_s= \lambda (\alpha_w - \frac{\pi}{2}).
\end{equation}
The polar equation of the jet shape in the reference system $(x',y')$ is given
by
\begin{equation}\label{rphip}
r(\phi')=r_w 
\left[\frac{\sin\alpha_w}{\sin(\alpha_w-\phi'/\lambda)}\right]^\lambda, \;\; 
\lambda(\alpha_w-\pi)\ < \phi' < \lambda\alpha_w.
\end{equation}
Therefore, the coordinates of the jet shape in the reference system $(x, 
y)$, centered on the jet source and with the $y$
axis along the tangent to jet/counterjet path, are given by
\begin{eqnarray}\label{xyphip}
x&=& \left[r(\phi')\cos\phi'-r_w\right]\sin\alpha_w-
 r(\phi')\sin\phi' \cos\alpha_w, \nonumber\\
y&=& \left[r(\phi')\cos\phi'-r_w\right]\cos\alpha_w+
 r(\phi')\sin\phi' \sin\alpha_w.
\end{eqnarray}
The jet shape described by Eqs.\ \ref{rphip} and \ref{xyphip} depends on
three parameters: 
the distance to the wind source, $r_w$;
the angle between the jet axis and the direction to the wind source,
$\alpha_w$; and 
the total deflection angle, $\psi_j$ (or $\lambda$, related through Eq.\ 
\ref{psi}). 

A least-squares fit of the three free parameters of the model was performed to
the positions of 43 knots, NA to NH from \citet{ang07}, knots A1 to 
I2 of the jet, and Z1 to N of the counterjet (present work), spanning a total
length along the jet axis of $\sim500''$.

\begin{figure}
\plotone{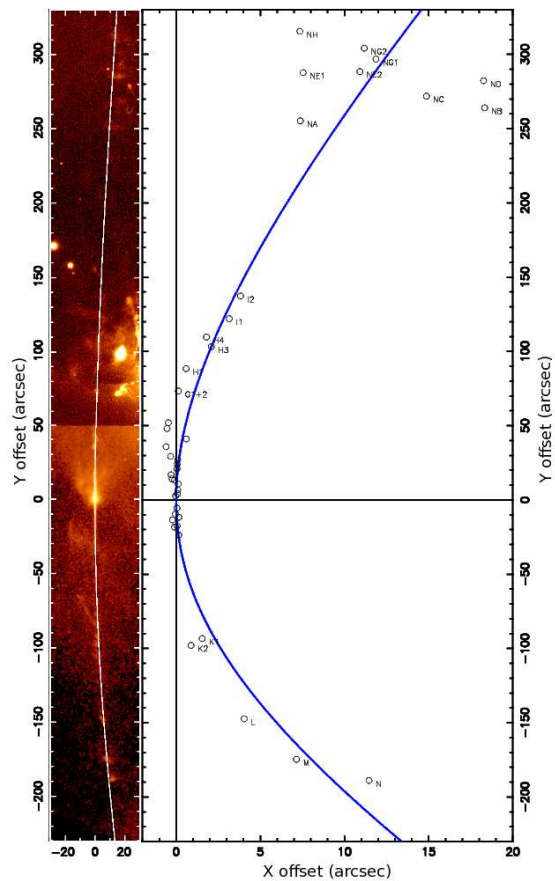}
\caption{Bending of the HH~30 jet and counterjet. 
\emph{Right:} The small circles mark the positions of the knots of the HH 30 jet
and counterjet, with the farther knots labeled. The continuum line shows the
best fit to the knot positions using the \citet{rag09a} model of jet bending by
an isotropic stellar side wind. The $x$ scale is amplified a factor of
$\sim10$.  
\emph{Left:} Best fit superposed on a [SII] image of the jet and counterjet.
The upper half-image is from 1999 \citep{ang07} and the lower
half-image is from 2010 (present work). The $x$ and $y$ scales are the same. 
}
\label{fig_bendfit}
\end{figure}

\begin{figure}
\plotone{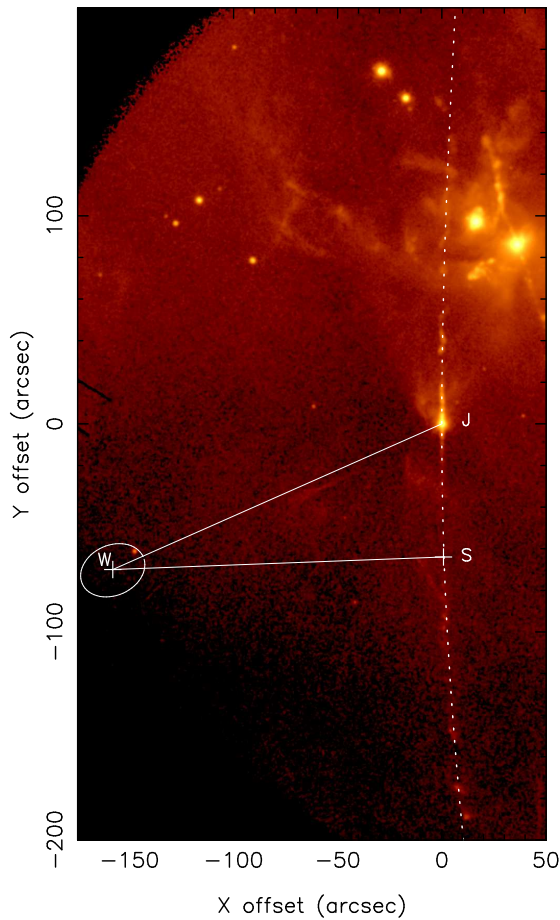}
\caption{Bending of the HH~30 jet and counterjet by an isotropic stellar wind.
The dotted line shows the best fit using the \citet{rag09a} model. The positions
of the HH 30 star ($J$), wind source ($W$) and stagnation point ($S$) are
indicated. The wind source is located at an offset position  $(-158'',-70'')$.
The ellipse shows the 1-$\sigma$ uncertainty (68\% confidence) in the position
of $W$. The stagnation point is located at $(+0\farcs8, -64'')$.
}
\label{fig_bendW}

\end{figure}

The best fit was obtained for
$r_w=173''\pm15''$ ($24200\pm2100$ AU in projection), 
$\alpha_w=66\degr\pm4\degr$, 
$\psi_j=14\fdg8\pm0\fdg7$ (corresponding to $\lambda=0.918\pm0.004$), 
with an rms fit residual in $x$ of
$\epsilon_\mathrm{fit}=2\farcs22$ 
(see Fig.\ \ref{fig_bendfit}).
The errors quoted were estimated as the increment in each parameter that
increases the rms fit residual a factor of $(1+{\chi^2}_3/n)^{1/2}$, 
where $n$ is the number of points fitted, and
${\chi^2}_3$ is the value of $\chi^2$ for 3 degrees of freedom (the number of
free parameters) and 68\% significance (1-$\sigma$ uncertainty), 
${\chi^2}_3=3.53$ \citep{lam76}.
For these values of the parameters we obtain that
the stagnation point is located at $(x,y)=(+0\farcs8, -64'')$ 
($S$ position in Fig.\ \ref{fig_bendW}).
The wind source is located at a position $(x,y)=(-158'', -70'')$
($W$ position in Fig.\ \ref{fig_bendW}). 
Within the 1-$\sigma$ uncertainty ellipse of the $W$ position we found a
reddened star, identified in the 2MASS catalog (J04314418+181047),
with coordinates
$\alpha (J2000)= 04^\mathrm{h}31^\mathrm{m}44\fs180$,
$\delta (J2000)= +18\degr10'14\farcs78$, whose colors
($J-H=1.33$, $H-K=0.81$) correspond to those of a Classical T Tauri star (CTTS)
in the $(J-H),(H-K)$ diagrams \citep{lad92, rom09}.

The total deflection of the jet, $\psi_j=14\fdg8\pm0\fdg7$, implies a value of
the parameter $\epsilon=11.2\pm0.6$. As stated in Eq.\ \ref{eq_epsilon},
$\epsilon$ is related to the physical properties of the jet and the wind
source. Taking a jet velocity $v_j=100$ km~s$^{-1}$, and an isothermal sound
speed of the jet $c_j=10$ km~s$^{-1}$, we obtain
\begin{equation}
\frac{\dot{M}_j v_j}{\dot{M}_w v_w}= 
\left(\frac{\epsilon}{2}\,\frac{c_j}{v_j}\right)^2= 0.31\pm0.02.
\end{equation}
The mass-loss rate of the HH 30 jet and counterjet has been estimated to be
$\dot{M}_j\simeq2.6\times10^{-9}$ $M_\odot$ yr$^{-1}$ \citep{bac99}.
For a jet velocity $v_j=100$ km~s$^{-1}$, the HH 30 momentum rate is
$\dot{M}_j v_j\simeq2.6\times10^{-7}$ $M_\odot$ yr$^{-1}$ km s$^{-1}$.
Thus, the wind source needed to deflect the jet has to have a momentum rate
\begin{equation}
\dot{M}_w v_w \simeq 
(8.3\pm0.5)\times10^{-7}~M_\odot\mathrm{~yr^{-1}~km~s^{-1}}.
\end{equation}
Mass-loss rates for CTTS in Taurus are of the order of 
$\dot{M}_\mathrm{w} \simeq 10^{-10}$--$10^{-8}$ $M_\odot$ yr$^{-1}$, and 
up to one order of magnitude higher for continuum CTTS \citep{har05}. Typical 
CTTS wind velocities are of the order of 100 km~s$^{-1}$.
Thus, for a CTTS we can expect
\begin{equation}
\dot{M}_w 
v_w\simeq10^{-8}\mathrm{-}10^{-6}~M_\odot\mathrm{~yr^{-1}~km~s^{-1}}.
\end{equation}
 This range of momentum rates encompasses the value needed to cause the 
observed deflection of the jet, given by Eq. 9. Thus, the star 
J04314418+181047 could be the wind source responsible for the deflection 
of the HH 30 jet, provided its mass-loss rate falls in the higher end 
of values found for CTTS.

If a CTTS can produce the observed bending of HH 30, this would imply that
a vast majority of jets should be bent, since low-mass protostars are abundant.
However, the bending of a jet is not easy to detect since it is only
noticeable when the jet is imaged over a long length. In the case of HH 30, the
images span a length of $500''$, or 0.35 pc.

\section{Jet wiggling}

Following the work of \citet{ang07}, we consider that the wiggling of the HH 30
jet and counterjet is a consequence of the presence of a companion star to the
jet source. \citet{ang07}, using the formulation given by \citet{mas02}, 
analyze the extreme cases where the dominant effect is either the orbital
motion of the jet source in a binary system, or the precession of the ejection
axis of the jet because of tidal interactions between the disk where the jet
originates and a companion star. Here we develop the work of
\citet{rag09b}, and we consider a physical system, in which both the orbital
motion and precession can be present simultaneously, to fit the HH 30 jet and
counterjet wiggling shape.

\subsection{Orbital and precession periods}

We consider a binary system with a circular orbit, being $m_j$ the mass of the
jet source, $m_c$ the mass of the companion, and $m=m_j+m_c$ the total mass of
the system. We will call $\mu_c$ the mass of the companion relative to the total
mass, so that 
\begin{eqnarray}\label{m1m2}
m_j&=&(1-\mu_c)\,m, \nonumber\\
m_c&=&\mu_c\,m.
\end{eqnarray}
Let $a$ be the binary separation (i.e.\ the radius of the relative orbit).
Therefore, the orbital radius of the jet source with respect to the binary's
center of mass (i.e.\ the radius of the jet source absolute orbit) is
\begin{equation}\label{mua}
r_o=\mu_c\,a,
\end{equation}
and the orbital velocity of the jet source is given by
\begin{equation}\label{eqvo}
 v_o={2\pi{}r_o\over\tau_o},
\end{equation}
where $\tau_o$ is the orbital period. The total mass of the binary system is
related to $\tau_o$, and $\mu_c$, $r_o$, or $a$, through Kepler's third law,
\begin{equation}\label{kepler}
\left(\frac{m}{M_\sun}\right)=
\mu_c^{-3}
\left(\frac{r_o}{\mathrm{AU}}\right)^3
\left(\frac{\tau_o}{\mathrm{yr}}\right)^{-2}=
\left(\frac{a}{\mathrm{AU}}\right)^3
\left(\frac{\tau_o}{\mathrm{yr}}\right)^{-2}.
\end{equation}

Let us consider that the disk of the jet source is tilted an angle $\beta$ with
respect to the orbital plane, and that it is precessing with a period $\tau_p$.
An approximate expression relating the orbital and precession periods can be
derived from Eq.\ 24 of \citet{ter98}, valid for a disk precessing as a rigid
body, by assuming that the disk surface density is uniform and that the rotation
is Keplerian,
\begin{equation}\label{terquem}
\frac{\tau_o}{\tau_p}=
\frac{15}{32}\,\frac{\mu_c}{(1-\mu_c)^{1/2}}\,{\sigma^{3/2}\cos\beta},
\end{equation}
where $\sigma=r_d/a$ is the ratio of disk radius to binary separation.
Since it is expected that the size of the disk is truncated by tidal
interaction with the companion star in such a way that $1/4\le\sigma\le1/2$
\citep{ter98}, we will adopt a value of $\sigma=1/3$. With this value of
$\sigma$, Eq.\ 15 gives:
$\tau_o/\tau_p=0.09(\mu_c/\sqrt{1-\mu_c}\,)\cos\beta$. Note that, except 
for values of $\mu_c$ near to 1, the precession period is much longer than the
orbital period. 

\begin{figure}
\plotone{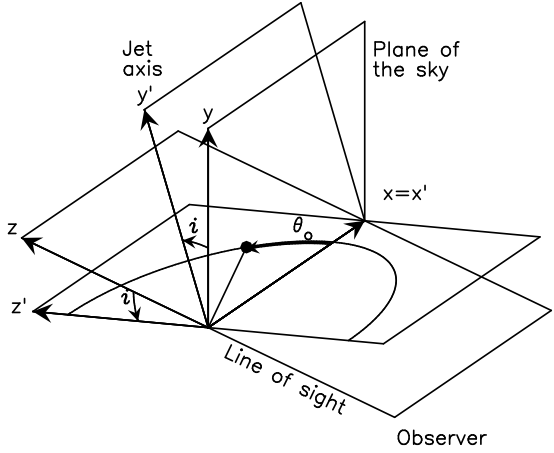}
\caption{Reference systems used for studying the jet and counterjet wiggling. 
The jet axis has an inclination $i$ with respect to the plane of the sky.
The orbit is contained in the $(x',y')$ plane. The orbital phase angle,
$\theta_o$, is measured counter-clockwise from the $x'$ axis.
}
\label{fig_coord}
\end{figure}

\subsection{Velocity and position of the knots as a function of time}

For describing the jet we use a (linear) coordinate system $(x',y',z')$, 
with origin in the jet source, and where $(x',z')$ is in the orbital 
plane, the $x'$-axis is the intersection of the orbital plane with the 
plane of the sky, and the $y'$-axis coincides with the orbital axis, at an 
inclination angle $i$ with respect to the plane of the sky (pointing away 
from the observer). The coordinate system is illustrated in Fig.\ 
\ref{fig_coord}.

As in \S \ref{pmotion}, for describing the observations we use an
(angular) coordinate system $(x,y,z)$, where $(x,y)$ is in the plane of the sky
(the $y$-axis in the jet direction and the $x$-axis perpendicular with values 
increasing from left to right), and the $z$-axis points away from the observer
(see Fig.\ \ref{fig_coord}).
The transformations between both systems of coordinates are
\begin{equation}\label{ccoord}
x= \frac{x'}{D},\hspace{1em}
y= \frac{y'\cos{i}-z'\sin{i}}{D},
\end{equation}
where $D$ is the distance from the source to the observer, which accounts for
the linear to angular transformation.

Let us assume that the velocity of the jet has a component
perpendicular to the disk plane, $v_j$, and a component in the orbital plane,
caused by the orbital motion of the jet source, $v_o$. The components of the 
velocity of a jet parcel, or knot, are
\begin{eqnarray}\label{vj}
v_{x'} &=& -v_o\sin\theta_o\pm v_j\sin\beta\cos\theta_p, \nonumber\\
v_{y'} &=& \pm v_j\cos\beta, \\
v_{z'} &=& v_o\cos\theta_o\pm v_j\sin\beta\sin\theta_p, \nonumber
\end{eqnarray}
with the sign of the $v_j$ term being positive for $y'>0$, and negative for
$y'<0$. The angles $\theta_o$ and $\theta_p$ are the orbital and precession
phase angles, measured from the $x'$ axis (perpendicular to the line of sight),
at the epoch of launch, $t_\mathrm{lch}$, when the jet parcel was ejected,
\begin{eqnarray}\label{theta}
\theta_o &=&  \frac{2\pi}{\tau_o}\,(t_\mathrm{lch}-t_0)+\varphi_o, \nonumber\\
\theta_p &=& -\frac{2\pi}{\tau_p}\,(t_\mathrm{lch}-t_0)+\varphi_p,
\end{eqnarray}
where  $\varphi_o$, $\varphi_p$ (between $0$ and $2\pi$) are the orbital and
precession phase angles at an arbitrary epoch of reference, $t_0$.  Note that we
assume that the precession is retrograde  \citep[see the discussion on
retrograde precession in binary systems;][]{mon09}, i.e.\ the orbital phase
angle $\theta_o$ increases with time, while the precession phase angle
$\theta_p$ decreases with time. The expression of Eq.\ \ref{theta} corresponds
to orbital phase angles increasing from $x'$ to $z'$ (see Fig.\
\ref{fig_coord}), i.e.\ a counter-clockwise rotation as seen from positive
values of $y'$ (and clockwise for precession).  In case of clockwise orbital
rotation (and counter-clockwise precession), the orbital phase angle would
decrease from $x'$ to $z'$, and the velocity $v_{z'}$ (and the $z'$ coordinate,
see below), would have the opposite sign. The difference is only noticeable when
the inclination angle $i$ is large, since $z'$ appears multiplied by $\sin{i}$
in the transformation given by Eq.\ \ref{ccoord}.

The velocity of a jet parcel, or a knot, in the plane of the sky can be found
from Eq.\ \ref{vj} by
using a coordinate transformation similar to that of Eq.\ \ref{ccoord},
\begin{eqnarray}\label{vxy}
v_x &=& -v_o\sin\theta_o\pm v_j\sin\beta\cos\theta_p,\nonumber\\
v_y &=& -v_o\cos\theta_o\sin{i} 
 \pm v_j(\cos\beta\cos{i}-\sin\beta\sin\theta_p\sin{i}).
\end{eqnarray}
The trajectory of this jet parcel, or knot, as a function of time is a straight
line given by
\begin{eqnarray}\label{xpypzp}
x' &=& r_o\cos\theta_o +
 (-v_o\sin\theta_o\pm v_j\sin\beta\cos\theta_p)\,(t-t_\mathrm{lch}),\nonumber\\
y' &=& \pm v_j\cos\beta\,(t-t_\mathrm{lch}), \\
z' &=& r_o\sin\theta_o +
 (v_o\cos\theta_o\pm v_j\sin\beta\sin\theta_p)\,(t-t_\mathrm{lch}). \nonumber
\end{eqnarray}
Note that the position of a knot at the epoch of its ejection
($t=t_\mathrm{lch}$) is in the orbital plane ($y'=0$). By using the coordinate
transformation of Eq.\ \ref{ccoord} we obtain that the position of the 
knot in the plane of the sky, as a function of time, is given by 
\begin{eqnarray}\label{xycoord}
x &=& \frac{1}{D} [r_o\cos\theta_o +
 (-v_o\sin\theta_o\pm v_j\sin\beta\cos\theta_p)\,(t-t_\mathrm{lch})],\nonumber\\
y &=& \frac{1}{D} [-r_o\sin\theta_o\sin{i}+(-v_o\cos\theta_o\sin{i} \\ 
&&{}
\pm v_j[\cos\beta\cos{i}-\sin\beta\sin\theta_p\sin{i}])\,(t-t_\mathrm{lch})].
\nonumber
\end{eqnarray}

\subsection{Constant velocity jet: shape of the jet}

If the jet velocity $v_j$ is constant, all the knots have the same velocity,
and the jet at a given epoch, $t=t_\mathrm{obs}$, has a shape that can be
obtained by eliminating the time dependence of Eq.\ \ref{xpypzp}, substituting 
$t-t_\mathrm{lch}=t_\mathrm{obs}-t_\mathrm{lch}=|y'|/(v_j\cos\beta)$ in the
equations. The orbital and precession phase angles can be expressed as
\begin{eqnarray}\label{thetaD}
\theta_o &=& \frac{2\pi}{\tau_o}
  \left(t_\mathrm{obs}-t_0-\frac{|y'|}{v_j\cos\beta}\right)+\varphi_o, 
  \nonumber\\
\theta_p &=& -\frac{2\pi}{\tau_p}
  \left(t_\mathrm{obs}-t_0-\frac{|y'|}{v_j\cos\beta}\right)+\varphi_p.
\end{eqnarray}
For simplicity, we can introduce the parameter $\alpha$ (which plays a role
similar to the angle of precession $\beta$),
\begin{equation}\label{alpha} 
\tan\alpha=\frac{v_o}{v_j\cos\beta}, 
\end{equation}
so that $v_o\,(t_\mathrm{obs}-t_\mathrm{lch})=|y'|\tan\alpha$. 
The angle $\alpha$ is the half-opening angle of the jet in the $(x',y')$ plane,
containing the jet axis, in the absence of precession. 
With these changes, the shape of the jet is given by
\begin{eqnarray}\label{shape}
x' &=& r_o\cos\theta_o-|y'|\tan\alpha \sin\theta_o+y'\tan\beta\cos\theta_p, 
 \nonumber\\
z' &=& r_o\sin\theta_o+|y'|\tan\alpha \cos\theta_o+y'\tan\beta\sin\theta_p.
\end{eqnarray} 
The shape of the jet in the plane of the sky $(x,y)$ is obtained from the last
equation, using the transformation of Eq.\ \ref{ccoord}.

\subsection{Variable velocity jet}

The proper motions measured for the HH 30 knots (\S \ref{pmotion}) show that the
ejection velocity is significantly different for some knots. 
Let us consider that $v_j$ (i.e.\ the ejection velocity component perpendicular
to the disk plane), instead of being the same for the full jet, can be different
for different parts of the jet. In this case, the jet has a shape composed of
different parts moving at different velocities. 
Let us estimate the $x$ position of each knot predicted by a model with given
values of the parameters of the  orbital motion 
($v_o$, $\tau_o$, $\varphi_o$),
and of those of the  precession 
($\beta$, $\tau_p$, $\varphi_p$). 

For each knot we know, from our observations, its position, $y$, and proper
motion, $\mu_y$, along the jet axis.
We can approximate the knot kinematic age by $t_\mathrm{kin}\simeq y/\mu_y$.
This approximation does not take into account the shift in position of the jet
source caused by its orbital motion, so it holds for $y\gg r_o/D$, or for small
inclination angles $i$. The epoch of launch of the knot is given by
$t_\mathrm{lch}=t_\mathrm{obs}-t_\mathrm{kin}$, and we can estimate the
orbital and precession phase angles at the epoch of launch, given by Eq.\
\ref{theta}. Once the orbital and precession phase angles are known, the 
ejection velocity component perpendicular to the disk plane for this knot,
$v_j$, can be calculated from Eq.\ \ref{vxy},
\begin{equation}\label{vv_v}
v_j= \frac{D\mu_y+v_o\cos\theta_o\sin{i}}
{\cos\beta\cos{i}-\sin\beta\sin\theta_p\sin{i}}.
\end{equation}
Therefore, the position of the knot in the plane of the sky at the epoch 
of observation ($t=t_\mathrm{obs}$) can be obtained from Eq.\ 
\ref{xycoord}, for $t-t_\mathrm{lch}=t_\mathrm{kin}$, 
\begin{eqnarray}\label{vvmodel}
x &\simeq& \frac{1}{D} [r_o\cos\theta_o +
 (-v_o\sin\theta_o
 +v_j\sin\beta\cos\theta_p)\,t_\mathrm{kin}], \nonumber\\
y &\simeq& \frac{1}{D} [-r_o\sin\theta_o\sin{i} + 
 (-v_o\cos\theta_o\sin{i} \\
 &&{}+v_j[\cos\beta\cos{i}
 -\sin\beta\sin\theta_p\sin{i}])\,t_\mathrm{kin}]= \nonumber\\
 &=& y-\frac{r_o}{D}\sin\theta_o\sin{i}. \nonumber
\end{eqnarray} 
The first equation gives the $x$ coordinate of the knot, and the second equation
shows that the $y$ coordinate of the knot is derived consistently by the model,
for $y\gg r_o/D$, or for small inclination angles $i$.

\subsection{Model parameters and limiting cases} 

In general, if the inclination angle $i$ is known, and both 
orbital motion and precession can be fitted to the data, the model depends 
on six parameters, which can be chosen to be 
the mass function, $m\mu_c^3$ (see below), 
the orbital period, $\tau_o$, 
the orbital phase angle at epoch $t_0$, $\varphi_o$, 
the precession angle, $\beta$, 
the precession period, $\tau_p$, and 
the precession phase angle at epoch $t_0$, $\varphi_p$. 
The rest of  parameters of the system can be obtained from these:
$\mu_c$ is derived from Eq.\ \ref{terquem};
$m_j=(1-\mu_c)m$, and $m_c=\mu_c m$ (Eq.\ \ref{m1m2});
$r_o$ and $a$ are derived from Eq.\ \ref{kepler}; and
$v_o$ is derived from $\tau_o$ and $r_o$ (Eq.\ \ref{eqvo}).

The ``pure'' orbital model of \S\ 4.1.1 of \citet{ang07} corresponds to 
$\beta=0$ in the previous equations ($\tau_p$, and $\varphi_p$ are undefined).
In the case of a constant velocity jet, the shape of the jet (Eq.\ \ref{shape})
has a plane symmetry (``C'' symmetry) with respect to the $x$ axis. For such a
``pure'' orbital model, $\alpha$ (Eq.\ \ref{alpha}) is the half-opening angle
of the jet cone, 
and the orbital period $\tau_o$ is determined by the jet velocity and the
wavelength of the jet wiggles in the plane of the sky. 
However the total mass of the system, $m$, and 
the mass of the companion relative to the total mass, $\mu_c$, 
cannot be determined independently, since the model depends only 
on the mass function $m\mu_c^3$ (Eq.\ \ref{kepler}). Thus, the ``pure'' 
orbital model depends on three parameters, which can be chosen to be 
the  mass function, $m\mu_c^3$, 
the orbital period, $\tau_o$, and 
the orbital phase angle at epoch $t_0$, $\varphi_o$. 
This will be also the case of the general model when precession is not relevant
in shaping the observed jet, for instance when the precession angle $\beta$ is
small compared to $\alpha$, or when the wavelength of the wiggles caused by
precession is longer that the length of the observed jet. 

The ``pure'' precession model of \S\ 4.1.2 of \citet{ang07} is obtained  from
the previous equations in the limiting case $\mu_c\rightarrow0$. For  this limit
the orbital radius $r_o$, the orbital velocity $v_o$, and  $\alpha$ tend to
zero, while the binary separation $a$ and orbital period  $\tau_o$ remain
finite. In this case the orbital terms in Eqs.\ \ref{shape} and \ref{vvmodel}
vanish, and, in the case of a constant velocity jet, the shape of the jet  has a
point symmetry (``S'' symmetry) with respect to the origin. However,  for a real
system ($\mu_c>0$), there is always orbital motion, even if  precession is
dominant. For such a precession-dominant model  ($\alpha\ll\beta$), the
precession angle $\beta$ is the half-opening angle  of the jet cone, and the
precession period $\tau_p$ is determined by the  jet velocity and the wavelength
of the jet wiggles in the  plane of the sky. Thus, three parameters, 
the precession angle, $\beta$, 
the precession period, $\tau_p$, and 
the precession phase angle at epoch $t_0$, $\varphi_p$, 
can be obtained from the fit. Then, for each value of  $\mu_c$ in the interval
$0<\mu_c<1$, an orbital period $\tau_o$ is  obtained from Eq.\ \ref{terquem}
and, since $\alpha\ll\beta$,  the following upper limits are found for the
binary  separation, $a$, (Eqs.\ \ref{mua}, \ref{eqvo}, \ref{terquem},
\ref{alpha})
\begin{equation}\label{amax}
a \ll \frac{0.09}{2\pi}\,\frac{1}{(1-\mu_c)^{1/2}}\,
\tau_p\,v_j\,\sin\beta\,\cos\beta,
\end{equation}
and for the mass of the jet source, $m_j$, (Eqs.\ \ref{m1m2}, \ref{kepler},
\ref{terquem}, \ref{alpha})
\begin{equation}\label{mjmax}
m_j \ll \frac{0.09}{8\pi^3}\,\frac{(1-\mu_c)^{1/2}}{\mu_c^2}\,
\tau_p\,v_j^3\,\sin^3\beta\,\cos\beta.
\end{equation}

\subsection{Fitting procedure}
\label{fitting}

For the fitting process we consider as known the inclination angle of the 
jet axis with respect to the plane of the sky, $i$. As discussed in 
\citet{lop96} and \citet{ang07} the inclination angle is small, and so, 
the model depends weakly on $i$. We have taken $i=5\degr$. Another 
parameter considered as known is the offset in the $y$ direction of the HH 
30 star position with respect to the intensity peak of HH 30 (knot A0). We 
have taken $-0\farcs51$, the same value as in \citet{ang07}.

The model of variable velocity jet was used to fit the $x$ positions of the
knots A to E of the jet  and the knots Z1 to Z6 of the counterjet in the 2010
image (Tables  \ref{pmjet} and \ref{pmcjet}). The knot D4 of the jet was
excluded from  the fit because it was identified only in the 2010 image and its
proper  motion could not be measured.

The fitting strategy was to sample the six-dimensional parameter space,
defined by the parameters $m\mu_c^3$, $\tau_o$, $\beta$, $\varphi_o$ and
$\varphi_p$.
Several sampling methods were tested, i.e.\ regular grid, random, and Halton
sequence  \citep{hal64}. We discarded the regular grid method because some
parameters (for instance, $\tau_o$) can have evenly spaced values that fit the
data. We adopted a Halton quasi-random sequence because it samples the space
parameter more evenly than a purely random sequence, and the convergence of the
fitting procedure to the minimum of the rms residual is faster. 

Several runs of the fitting procedure were performed using different  number of
sample points (typically between $10^6$ and $10^7$), and  different ranges of
the parameters. The ``pure'' orbital model, depending three parameters, 
$m\mu_c^3$, $\tau_o$, and $\varphi_o$was also  fitted, by adopting a fixed value
$\beta=0$. 

Once a minimum of the rms fit residual was found, the uncertainty in the
parameters fitted was found as the increment of each of the six (or three)
parameters of the fit necessary to increase the rms fit residual a factor of
$(1+{\chi^2}_p/n)^{1/2}$, where 
$n$ is the number of knots whose position was fitted, 
$p$ is the number of parameters fitted, and 
${\chi^2}_p$ is the value of $\chi^2$ for $p$ degrees of freedom 
(the number of free parameters) and 68\% significance (1-$\sigma$ uncertainty),
i.e.\ ${\chi^2}_3=3.53$, ${\chi^2}_6=7.04$ \citep{lam76}.

\subsection{Results of the model fitting}
\label{results}

Let us first consider a system where precession and orbital motion are  present
in the jet, and both contribute to determine the shape of the jet. This
condition means that the half-opening angles due to orbital motion, $\alpha$,
and to precession, $\beta$, are of the same order. For the range of masses
considered for the HH 30 jet source and the companion (0.06--1 $M_\odot$; see
below), the orbital period is substantially shorter than the precession period. 
Thus, the jet shapes obtained from the model present a short-scale wiggling
corresponding to the orbital period $\tau_o$, and a large-scale wiggling
corresponding to the precession period $\tau_p$ \citep[see, for instance, Fig.\
3 of][]{rag09b}. Although such a kind of model can reproduce the observed $x$
positions of the limited number of knots included in the fitting, it predicts a
small-scale wiggling shape between these knots, which is not observed in the HH
30 jet images. Thus, the model that fits the data has to be either
precession-dominant ($\alpha\ll\beta$), or orbital-motion-dominant
($\beta\ll\alpha$).

The precession-dominant model can be excluded from simple
considerations.  Since the observed wiggles of the jet in the plane of the sky
have a  wavelength of $\sim 16''$ \citep[][see also Fig.\ 14]{ang07}, a jet 
velocity of $v_j\simeq 100$ km s$^{-1}$ would result in $\tau_p\simeq100$ yr,
assuming a precession-dominant scenario.  Since the  observed half-opening angle
of the jet cone is $\sim1\fdg5$ \citep{ang07}, and $\alpha\ll\beta$, the
precession angle should be $\beta\simeq1\fdg5$.
We know that the mass of the jet  source cannot be higher than $\sim1$
$M_\odot$. Higher values of the  mass are unlikely since the bolometric
luminosity of the system is less  than $1~L_\odot$ \citep{cot01}. It can neither
be lower than the mass of a brown dwarf,  $\sim0.06$ $M_\odot$. 
Using these constraints and Eqs.\
\ref{amax}, \ref{mjmax}, we can  derive an upper limit for the binary separation
for the allowed range of  masses of the jet source. The value obtained for the
binary separation in  all cases is $a\ll1$ AU. This is extremely improbable
since the radius of  the accretion disk should be even smaller, while typical
sizes of  accretion disks are more than one order of magnitude larger. Is is
worth  noting that \citet{ang07} obtain higher values of the proper  motion
velocities ($\sim 200$ km s$^{-1}$), resulting in a smaller value  for the
estimated period, $\tau_p\simeq53$ yr, that did not allow them to  discard a
precession-dominant scenario for the allowed range of masses.
Thus, the wiggles observed in the plane of the sky have to be mainly 
caused by the orbital motion of the jet source. Precession, if any, is not 
important for shaping the jet and counterjet.

\begin{deluxetable*}{lllll}
\tablewidth{0pt}
\tablecaption{%
Results obtained from the fit to the jet and counterjet knots.\tablenotemark{a}
}
\tablecolumns{5}
\tablehead{
\colhead{Parameter} & \colhead{(Units)} 
&\colhead{Value} &\colhead{Value} &\colhead{Value}
}
\startdata
Knots fitted                &              & ABCE\tablenotemark{b}& ABCDE\tablenotemark{b}& ABCDEZ\tablenotemark{b}\\
Initial rms                 & (arcsec)     & $0.360$              & $0.321$               & $0.272$                \\
Fit rms                     & (arcsec)     & $0.090$              & $0.087$               & $0.104$                \\
$m\mu_c^3$                  & ($M_\odot$)  & $0.016\pm0.006$      & $0.015\pm0.009$       & $0.014\pm0.006$        \\
$\tau_o$                    & (yr)         & $110.1\pm1.8$        & $110.6\pm1.3$         & $113.9\pm1.7$          \\
$\varphi_o$\tablenotemark{c}& (rad)        & $2.23\pm0.17$        & $2.15\pm0.17$         & $1.65\pm0.20$          \\
$r_o$                       & (AU)         & $5.8\pm0.7$          & $5.7\pm1.1$           & $5.7\pm0.9$            \\
$v_o$                       & (km s$^{-1}$)& $1.6\pm0.2$          & $1.5\pm0.3$           & $1.5\pm0.2$            \\
$\beta$\tablenotemark    {d}& (deg)        & 0                    & 0                     & 0                      \\
$m_j$\tablenotemark      {e}& ($M_\odot$)  & 0.06--1.0            & 0.06--1.0             & 0.06--1.0              \\
$\mu_c$\tablenotemark    {f}&              & 0.51--0.23           & 0.50--0.23            & 0.49--0.22             \\
$m_c$\tablenotemark      {f}& ($M_\odot$)  & 0.06--0.30           & 0.06--0.29            & 0.06--0.29             \\
$a$\tablenotemark        {f}& (AU)         & 11.4--25.1           & 11.4--25.1            & 11.5--25.6             
\enddata
\tablenotetext{a}{Obtained for a distance $D=140$ pc 
and assuming an inclination angle of the jet axis
with respect to the plane of the sky $i=5\degr$.}
\tablenotetext{b}{Knot velocities used:
$v_j=98$ km s$^{-1}$ (A, B, C, E, Z4); 
$v_j=181$ km s$^{-1}$ (Z1, Z3, Z5);  
$v_j=234$ km s$^{-1}$ (D); 
$v_j=307$ km s$^{-1}$ (Z2, Z6).}
\tablenotetext{c}{At the epoch $t_0=2010$ Dec 01.} 
\tablenotetext{d}{Adopted (see text).} 
\tablenotetext{e}{Range of masses considered for the jet source (see text).} 
\tablenotetext{f}{The first value is for the minimum value of $m_j$, 
and the second one is for the maximum.} 
\label{tmodel2}
\end{deluxetable*}

\begin{figure}
\plotone{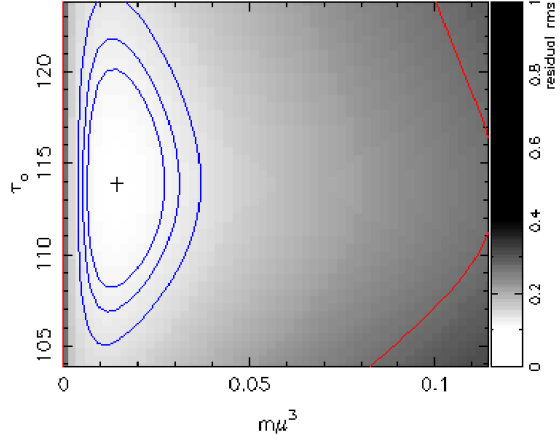}
\caption{Rms fit residual of the ``pure'' orbital model fitted to the HH 30
knots of the jet and counterjet, as a function of mass function $m\mu_c^3$ and
orbital period $\tau_o$. The cross marks the position of the best fit, for which
the rms is $0\farcs104$.
The blue contours indicate the confidence levels 
68\% (1-$\sigma$ uncertainty),
90\%, and 
99\%,
while the red contour indicates the rms $x$ position of the knots with respect
to the jet axis, 
$0\farcs272$.
}
\label{frmsmodel}
\end{figure}

In order to fit the shape of the jet and counterjet, we proceeded in  various
steps. For each step we used the procedure described in \S \ref{fitting}. First
we fitted the ``pure'' orbital model to the knots of the jet that move at a
similar (low) velocity, and where the wiggling is  more evident: knots A, B, C,
and E. For these knots we adopted the same  velocity, the error-weighted average
of their velocities, 98 km s$^{-1}$.  The result of this first fit is shown in
Table \ref{tmodel2}.

The next step was to include in the fit the rest of knots of the jet, the
high-velocity knots D. For these knots, we also adopted as velocity the
error-weighted average of their velocities, 234 km s$^{-1}$ (see Fig.\
\ref{fvyy}). The same values (within errors) of the parameters obtained for the
first fit, were able to fit all the knots of the jet (see Table \ref{tmodel2}).

The third step was to include in the fit the knots of the counterjet. As stated
before, the knots of the counterjet have very different velocities. We
classified the counterjet knots into three groups. The medium-velocity group is
formed by knots Z1, Z3, Z5a, and Z5b. Their error-weighted average velocity is
181 km s$^{-1}$. The low-velocity group contains only one knot, Z4, for which we
adopted the same velocity as the low-velocity knots of the jet, 98 km s$^{-1}$.
Finally, the extremely-high velocity group, with knots Z2 and Z6, for which we
adopted the error-weighted average of their velocities, 307 km s$^{-1}$ (see Fig.\
\ref{fvyy}). Once again, the same values (within errors) of the parameters
obtained for the first fit, were able to fit all the knots of the jet and
counterjet (see Table \ref{tmodel2}). 

The last step was to include precession in the model, as a perturbation of the
orbital motion (i.e.\ for $\beta\ll\alpha$). However, for this last fit the
parameters of the precession were not well constrained by the data, and the
rms fit residual did not improve with the inclusion of precession.

\begin{figure*}
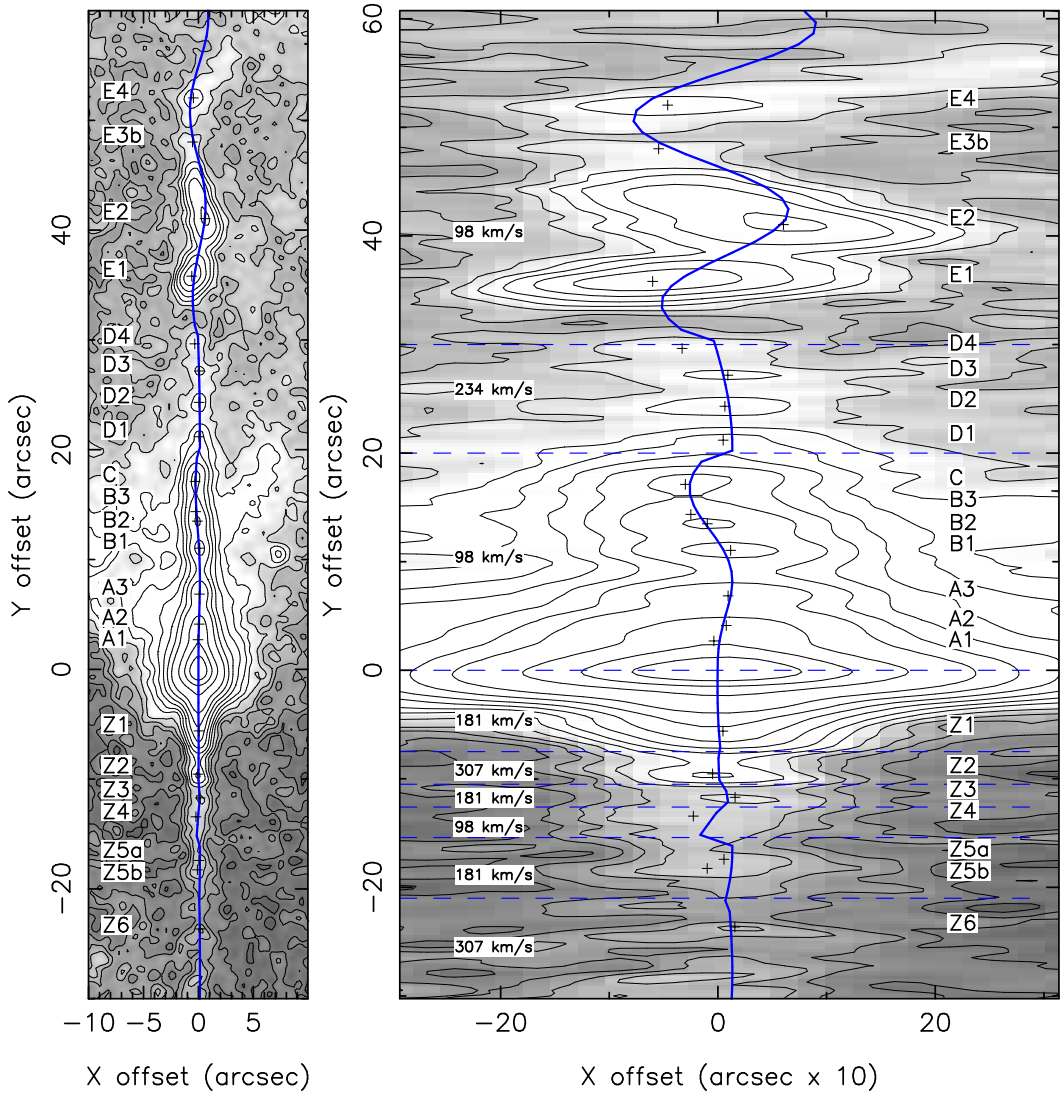

\center
\includegraphics[scale=0.6,clip=true]{f10a.eps}
\includegraphics[scale=0.6,clip=true]{f10b.eps}
\caption{2010 [SII] narrow-band image of the jet and counterjet of HH~30,
between $-30''$ and $+60''$ of the HH~30 star. 
\emph{Left:} Same scale in $x$ and $y$.
\emph{Right:} $x$ scale amplified a factor of 10. 
The measured positions of the knots are marked with crosses. 
The rms $x$ position of these knots with respect to the jet axis is
$0\farcs272$.
The continuum blue line is the jet shape calculated from the ``pure'' orbital
model  with a mass function $m\mu_c^3=0.014\pm0.006$ $M_\odot$
an orbital period $\tau_o=114\pm2$ yr,
and an orbital phase angle $\varphi_o=1.65\pm0.20$ rad at $t_0=2010$ Dec 01.
The rms fit residual in $x$ is $0\farcs119$.
The horizontal dashed lines mark the boundaries of the regions with different
knot velocities. The knot velocities are indicated for each region.
}
\label{fknotsused}
\end{figure*}

Thus, we consider that the best fit to the shape of the jet and counterjet 
is an orbital model, with a mass function $m\mu_c^3=0.014\pm0.006$ 
$M_\odot$, an orbital period $\tau_o=114\pm2$ yr, and an orbital 
phase angle $\varphi_o=1.65\pm0.20$ rad at $t_0=2010$ Dec 01. 
In Fig.\ \ref{frmsmodel} we 
show the rms fit residual as a function of the mass function $m\mu_c^3$ 
and orbital period $\tau_o$, around the best-fit values. The shape of the 
jet obtained from this model is shown in Fig.\ \ref{fknotsused}, where the 
continuum line indicating the shape of the jet has been obtained by 
interpolating the velocities adopted for the knots with a boxcar function, 
as shown in Fig.\ \ref{fvyy}. 

From the fitted parameters we obtain a radius of the absolute orbit 
$r_o=5.7\pm0.9$ AU, and an orbital velocity $v_o= 1.5\pm0.2$ km s$^{-1}$. The
orbital velocity obtained is low, similar to that obtained for other jets in
young stellar objects. For instance, in HH 111, $v_o= 3.1\pm1.9$ km s$^{-1}$
\citep{nor11}, and in HH 211, $v_o= 1.6\pm0.6$ km s$^{-1}$ \citep{lee10}.

\begin{deluxetable}{llllll}
\tablewidth{\columnwidth}
\tabletypesize{\small}
\tablecaption{%
Favored set of parameters of the HH 30 binary system\tablenotemark{a}. 
}
\tablecolumns{6}
\tablehead{
\colhead{${\tau_o}$\tablenotemark{b}}   &  
\colhead{${v_o}$\tablenotemark{c}}      &  
\colhead{${\varphi_o}$\tablenotemark{d}}& 
\colhead{${m_j}$\tablenotemark{e,f}}    &  
\colhead{${m_c}$\tablenotemark{e,g}}    &  
\colhead{$a$\tablenotemark{e,h}}        \\  
\colhead{(yr)}                          &
\colhead{(km s$^{-1}$)}                 &  
\colhead{(deg)}                         &
\colhead{($M_\odot$)}                   &  
\colhead{($M_\odot$)}                   &  
\colhead{(AU)}                            
}
\startdata
$114\pm2$ & $1.5\pm0.2$ & $95\pm11$ & $0.31\pm0.04$ & $0.14\pm0.03$& $18.0\pm0.6$
\enddata
\tablenotetext{a}{Obtained for a distance $D=140$ pc 
and assuming an inclination angle of the jet axis
with respect to the plane of the sky $i=5\degr$.}
\tablenotetext{b}{Orbital period.}
\tablenotetext{c}{Orbital velocity.}
\tablenotetext{d}{Orbital phase angle with respect to an axis 
perpendicular to the line of sight, at the epoch 2010 Dec 01.}
\tablenotetext{e}{Assuming a mass of the system $m=0.45\pm0.04$ $M_\odot$
\citep{pet06}.}
\tablenotetext{f}{Mass of the jet source.}
\tablenotetext{g}{Mass of the companion star.}
\tablenotetext{h}{Binary separation.}
\label{tbinary}
\end{deluxetable}

The values of the mass of the jet source and companion are not constrained by
the fit. For the range of possible masses of the jet source $m_j=0.06$--1
$M_\odot$ (see above), we obtain that the companion  has a range of masses
$m_c=0.06$--0.29 $M_\odot$, and the binary separation (radius of the relative
orbit) is in the range $a=11.5$--25.6 AU. 

\citet{pet06} find evidence of Keplerian rotation of the circumbinary disk in
their CO observations with the PdBI, and derive a value of $m=0.45\pm0.04$
$M_\odot$ for the central stellar mass. Using this value of $m$ and the
parameters of the fit, we obtain
a mass $m_j=0.31\pm0.04$ $M_\odot$ for the jet source, 
a mass $m_c=0.14\pm0.03$ $M_\odot$ for the companion, and 
a binary separation $a=18.0\pm0.6$ AU.

\section{Discussion}

\citet{ang07} carry out a detailed study of the morphology and proper motions
of the HH 30 jet. They find that the proper motion velocities of the jet knots
were of the order of 200 km s$^{-1}$. They also find that the jet path shows a
wiggling with a periodicity in the plane of the sky of $16''$ that translates
into a period of 53 yr using the derived proper motion velocity. It is shown
that this period and the observed shape of the jet can be explained either by
the orbital motion of the jet source or by precession of the jet axis because
of tidal interactions of a companion star. In the first case, the binary 
separation would be 9-18 AU, while in the precession scenario the binary
separation would be $<1$ AU. In either of the two scenarios the binary 
separation is much smaller than the radius ($\sim250$ AU) of the disk observed
with the HST \citep{bur96,sta99}, showing that it is a circumbinary rather than
a circumstellar disk, and that it cannot play a role in the jet collimation.
The true circumstellar disk associated with the HH 30 jet should be found at
scales of a fraction of the binary separation. In case of wiggling produced by
orbital motion, the jet/counterjet system is expected to present plane (``C'')
symmetry with respect to the equatorial plane, while in the case of precession
the jet/counterjet system is expected to present point symmetry with respect to
the position of the driving source. Since the images obtained by \citet{ang07}
only cover a small portion of the counterjet, they cannot discriminate 
between the two alternative scenarios.

In this paper we presented new observations that allowed us to image and 
measure proper motions both in the jet and in the counterjet. In principle,
these new data should be able to distinguish between the orbital and precession
scenarios. For instance, in HH 111 \citep{nor11} and HH 211 \citep{lee10} 
plane symmetries in the jet/counterjet indicative of orbital motion of the
driving source have been recently found. 
Unfortunately, the counterjet of HH 30 is weak, and appears less ordered than
the jet, with larger variations in the velocity of the knots. So, it is not
expected to show a clear symmetry as a constant velocity wiggling jet would do.

Interestingly, our new observations, covering a longer time span, allowed us to
derive more accurate proper motions and reveal that the velocities of the knots
are smaller than the values estimated by \citet{ang07}. The new values of the
proper motions imply that the wiggling period is about twice the previous value
(114 yr instead of 53 yr), allowing us to discard a significant contribution
of precession to the observed wiggling of the jet (see \S \ref{results}). 

Thus, based on our new observations we conclude that the observed shape of the
jet is a consequence of the orbital motion of the jet source in a binary system.
From our new results, and assuming that the central stellar mass is $m=0.45 \pm
0.04~M_\odot$, as estimated by \citet{pet06} from the CO kinematics of the disk,
we obtain that the binary separation (i.e.\ the radius of the relative orbit)
should be $18.0\pm0.6$ AU. Interestingly, \citet{gui08} observe a hole of
$37\pm4$ AU of radius in the CO and dust emission from the disk associated with
HH 30. These authors interpret this hole as produced by tidal effects in a
binary system, inferring a binary separation of $18\pm2$ AU, in excellent
agreement with our result. Therefore, these results give strong support to the
binary interpretation, and in particular to the orbital motion scenario we 
proposed for HH 30.

A binary separation of 18 AU corresponds to an angular separation of $0\farcs13$
at a distance of 140 pc. Thus, the proposed binary system associated with
the HH 30 jet can be angularly resolved by the new, expanded Jansky Very Large
Array (JVLA) and by the Atacama Large Millimeter/submillimeter Array (ALMA).
Unfortunately, according to our modeling of the orbital motion,  at the current
epoch ($\sim2012.5$) the two stars are roughly aligned along the line of sight
($\theta_o=100^\circ$), resulting in a projected angular separation of
$0\farcs02$, which is too small to be angularly resolved by the currently
available instrumentation. 
We should wait four years until the projected angular separation will reach
$\gtrsim0\farcs05$ and the presence of the proposed binary system could be
confirmed by a direct observation of a double source and/or to measure orbital
proper motions, as has been done  previously in other binary young stellar
objects \citep[e.g.\ L1551-IRS5;][]{rod03}

When the HST observations revealed the silhouette of a flared edge-on disk with
a radius of $\sim250$ AU perpendicular to the HH 30 jet, this system was
considered the archetype of a jet/circumstellar accretion disk system in a young
stellar object. However, the discovery that the central source is in fact a
binary, and therefore, that the disk is circumbinary rather than
circumstellar, has changed our understanding of this object. Tidal
truncation of the disk in a binary system suggests that the radius of the
circumstellar disk should be about $\sim 1/3$ of the binary separation
\citep[e.g.][]{ter99}. Thus, we expect that the ``true'' circumstellar accretion
disk associated with the driving source of the HH 30 jet must have a radius
$\la6$ AU. Therefore, the search for this circumstellar disk should be carried
out at very small scales ($\sim0\farcs05$)

Another interesting characteristic of the HH 30 system is that the jet and 
counterjet show clear differences. The knots of the jet and of the counterjet
cannot be grouped in pairs with similar separation from the star and similar
velocity. HH 30 seems to be, in this aspect, completely different, for instance,
from the jet and counterjet of HH 34, in which the jet and counterjet  show a
remarkable symmetry \citep{rag11}. Clear jet/counterjet asymmetries in the
physical properties have been found in several well studied jets, such as DG Tau
B \citep{pod11}, FS Tau B \citep{liu12}, DG Tau \citep{agr11}, or HH 30 itself
\citep{bac99}. Actually, strong velocity asymmetries between the red and blue 
lobes are frequent. \citet{hir94} note that about half of bipolar jets present a
strong asymmetry of about a factor two in velocity between the two lobes. In HH
30 we found that the average proper motion velocities in the counterjet are a
factor of 1.7 higher than in the jet. We also found a higher velocity dispersion
in the counterjet (see \S \ref{pmotion}).  In the frame of stellar wind models,
velocity asymmetries between the jet and counterjet have been attributed to
differences in pressure. In the frame of magneto-hydrodynamical disk winds, jet
asymmetries require an asymmetry in magnetic lever arms or in launch radii
between either sides of the disk, a situation that may occur naturally in an
asymmetric ambient medium. For example, if the ambient radiation field is
stronger on one side, this can increase the level of ionization on the surface
of the disk on that side, leading to enhanced mass load on the magnetic field
lines (smaller magnetic lever arm), and/or to a larger jet launching region
\citep[see][]{fer06}.

However, HH 30 also shows differences in velocity in the same lobe of the 
outflow. For the jet, there are at least two velocities, a low velocity of 
$\sim100$ km s$^{-1}$ for knots A, B, C, and E, and a high velocity of $\sim 
240$ km s$^{-1}$ for knots D. For the counterjet we find 
knots close to each other with different velocities, and there are at least
three different velocities, a low velocity of $\sim 100$ km s$^{-1}$ for knot
Z4, a medium velocity of $\sim 180$ km s$^{-1}$ for knots Z1, Z3, Z5a and Z5b,
and an extremely-high velocity of $>300$ km s$^{-1}$ for knots Z2 and Z6 (see
\S \ref{pmotion}). 
Indeed, we found that the knots Z1 and Z2 (and perhaps also Z5 and
Z6) of the counterjet appear to be launched nearly simultaneously with very
different velocities. These strong differences in velocity for knots located in
the same lobe of the outflow cannot be attributed only to asymmetries in the
ambient medium between the two sides of the disk. Therefore, launching from
different radii in an extended disk wind scenario provides the simplest
explanation for the different observed velocities. Following \citet{bla82}, the
asymptotic jet velocity as a function of the launch radius in the disk, $r_l$,
is expected to be of the order of 
$\sim100(M_\star/0.5~M_\odot)^{1/2}(r_l/1~{\rm AU})^{-1/2}$ km s$^{-1}$ 
for a value of the magnetic lever arm parameter of $\sim10$ 
\citep[i.e.\ ``the extended warm disk wind scenario with moderate lever arms''
described, e.g.\ in][and references therein]{fer06,pan12}. 
Assuming that the jet source has a mass of $0.31$ $M_\odot$ (see above), we
obtain that a range of launching radii $r_l=0.07$--0.6 AU is required to
explain the velocity range of $\sim100$--300 km s$^{-1}$ observed in the HH
30 jet/counterjet. This range of launching radii falls within the range of
values determined for T Tauri jets \citep[e.g.][and references
therein]{and03,pes04,fer06}, which range from 0.07 AU (the typical disk 
corotation radius) up to $\sim3$ AU.

\section{Conclusions}

Using data from observations at six different epochs, we have measured the 
proper motions and studied the morphology of the HH 30 jet/counterjet system.
Our main conclusions are summarized in the following.

\begin{itemize}

\item The motion of most knots is essentially ballistic, but the
jet/counterjet shows a large scale bending. This bending can be produced by a
relative motion of the HH 30 star with respect to its surrounding environment of
$\sim2$ km s$^{-1}$, caused either by a possible proper motion of the HH 30
star, or by the entraining of environment gas by the red lobe of the nearby
L1551-IRS 5 outflow. Alternatively, the bending can be produced by the stellar
wind from a nearby CTTS, identified in the 2MASS catalog as J04314418+181047. 

\item The average velocity of the knots of the jet is about twice than that of 
the counterjet. Velocity differences are also found in the same outflow lobe.
Most of the knots of the jet move at velocities of $\sim100$ km s$^{-1}$, but
several knots (knots D) move faster, at $\sim240$ km s$^{-1}$. The velocities of
the counterjet knots are less ordered, with knots moving, at least, at three
different velocities of about 100, 180, and $\ga300$ km s$^{-1}$. In particular,
we identified at least a pair of knots in the counterjet (knots Z1 and Z2) that
apparently have been launched simultaneously with very different velocities.

\item The asymmetry in the average velocities of the jet and counterjet may be
due to an asymmetric ambient medium. The observed differences in velocity of
the knots of the same outflow lobe can be attributed to different launching
radii in an extended ``warm'' disk wind scenario. The range of launching radii
in the circumstellar accretion disk required to account for the observed range
of velocities of the knots is $\sim0.07$ AU to $\sim0.6$ AU, which falls
within the range of typical values determined for T Tauri jets.

\item The observed wiggling of the HH 30 jet and counterjet is produced by the
orbital motion of the jet source in a binary system. Precession of the 
accretion disk, if present at all, appears to be a minor contribution in 
shaping the jet. The best fit to the shape of the jet is obtained with a binary
system with a mass function 
$m\mu_c^3=0.014\pm0.006$ $M_\odot$, an orbital period 
$\tau_o=114\pm2$ yr, and an orbital velocity of 
$v_o=1.5\pm0.2$ km~s$^{-1}$. Adopting a value of the total stellar mass of 
$m=0.45\pm0.04$ $M_\odot$ \citep{pet06}, 
we obtain that the mass of the jet driving source is 
$m=0.31\pm0.04$ $M_\odot$, the mass of the companion is 
$m=0.14\pm0.03$ $M_\odot$, and the binary separation is 
$18\pm0.6$ AU. 
The same value of the binary separation is inferred from the size of the inner
cavity observed in the circumbinary disk \citep{gui08}, assuming it is produced
by tidal truncation of the disk in a binary system. This result strongly
supports our binary interpretation first proposed by \citet{ang07}.

\end{itemize}

In summary, the HH 30 jet/counterjet system appears to originate likely from a
binary system of two stars of masses 0.31 $M_\odot$ (the jet driving source) and
0.14 $M_\odot$, separated 18 AU and orbiting with a period of 114 yr. The system
is surrounded by a circumbinary disk with an outer radius of $\sim 250$ AU and
with an inner radius $\sim 40$ AU. The driving source of the jet/counterjet
appears to be surrounded by a circumstellar accretion disk with an outer radius
$\la 6$ AU, and probably with an inner radius of $\sim 0.07$ AU. The observed
jet/counterjet knots reach velocities in the range $\sim100$--300 km s$^{-1}$,
arising from disk radii ranging from 0.07 AU to 0.6 AU. The two components of
the binary system are currently roughly aligned along the line of sight, but
their angular separation in the plane of the sky and orbital motions could be
determined by future observations with the JVLA or ALMA

\acknowledgments
G. A., R. E., R. L., A. R., and C. C.-G. are partially supported by Spanish MCI
grants AYA2008-06189-C03 and AYA2011-30228-C03, and FEDER funds.
We acknowledge Pau Estalella for his helpful comments on the use of the Halton
sequence. 
The data presented here were taken at
the 2.5 m Isaac Newton Telescope,
the 4.2 m William Herschel Telescope, and 
the 2.6 m Nordic Optical Telescope at 
the Observatorio del Roque de los Muchachos of 
the Instituto de Astrof\'{\i}sica de Canarias.

\clearpage

\end{document}